\documentclass[]{interact}

\usepackage{epstopdf}
\usepackage[caption=false]{subfig}

\usepackage[numbers,sort&compress]{natbib}

\bibpunct[, ]{[}{]}{,}{n}{,}{,}
\renewcommand\bibfont{\fontsize{10}{12}\selectfont}
\usepackage{subcaption}
\theoremstyle{plain}
\newtheorem{theorem}{Theorem}[section]

\theoremstyle{definition}

\theoremstyle{remark}

\usepackage{mathtools}
\DeclarePairedDelimiter{\ceil}{\lceil}{\rceil}

\usepackage{hyperref}

\usepackage[left]{lineno}

\newcommand{\answer}[1]{#1}
\usepackage{xcolor}
\usepackage{amsthm}
\newtheorem{mydefinition}{Definition}

\newcommand{\answerR}[1]{#1}

\newcommand{\answerRR}[1]{#1}

\newcommand{\answerRRR}[1]{#1}

\newcommand{\answereditor}[1]{#1}
\newcommand{\answerRrR}[1]{#1}
\begin{document}


\title{New flexible versions of extended generalized Pareto distribution for count data}

\author{
\name{Touqeer Ahmad \textsuperscript{a}\thanks{CONTACT Touqeer Ahmad. Email: touqeer.ahmad@ensai.fr} and Irshad Ahmad Arshad\textsuperscript{b}}
\affil{\textsuperscript{a}CREST, ENSAI, University of Rennes, France; \textsuperscript{b}Department of Statistics, Faculty of Sciences, Allama Iqbal Open University, H-8/4, Islamabad, 44000, Pakistan.}
}

\maketitle

\begin{abstract}
Accurate modeling is essential in integer-valued real phenomena, including the distribution of entire data, zero-inflated (ZI) data, and discrete exceedances. 
The Poisson and Negative Binomial distributions, along with their ZI variants, are considered suitable for modeling the entire data distribution, but they fail to capture the heavy tail behavior effectively alongside the bulk of the distribution. In contrast, the discrete generalized Pareto distribution (DGPD) is preferred for high threshold exceedances, but it becomes less effective for low threshold exceedances. However, in some applications, the selection of a suitable high threshold is challenging, and the asymptotic conditions required for using DGPD are not always met. To address these limitations, extended versions of DGPD are proposed. These extensions are designed to model one of three scenarios: first, the entire distribution of the data, including both bulk and tail and bypassing the threshold selection step; second, the entire distribution along with ZI; and third, the tail of the distribution for low threshold exceedances. The proposed extensions offer improved estimates across all three scenarios compared to existing models, providing more accurate and reliable results in simulation studies and real data applications.
\end{abstract}


\section{Introduction}\label{sec:1}

\begin{figure}
\centering
\subfloat[]{%
\resizebox*{7cm}{!}{\includegraphics{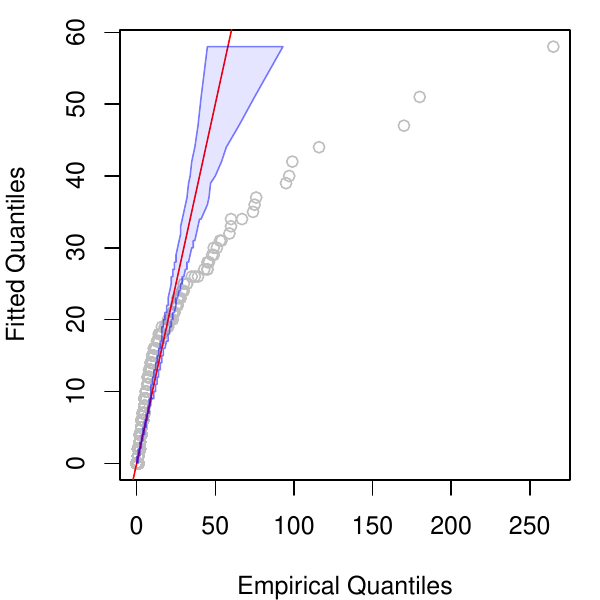}}}\hspace{5pt}
\subfloat[]{%
\resizebox*{7cm}{!}{\includegraphics{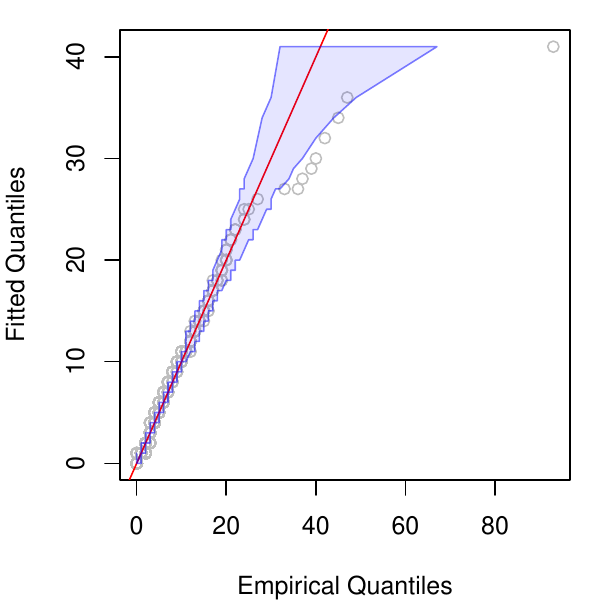}}}
\caption{\answereditor{The Q-Q plots of fitted ZINB to the count data sets. Upheld complaints of the automobile
insurance companies rankings of New York City (\textit{left}), and doctors visits to the hospital (\textit{right}).} } \label{fig:existmodel}
\end{figure}

Many statistical models are available for analyzing non-negative count data. Standard frameworks include the Poisson distribution, which suits count data with equal mean and variance, and the Negative Binomial distribution, which addresses overdispersion by allowing variance to exceed the mean. For data with an excessive number of zeros, ZI models such as the ZI Poisson (ZIP) and ZI Negative Binomial (ZINB) extend these distributions to account for additional zero inflation. The DGPD, on the other hand, is tailored for modeling discrete exceedances above a high threshold, making it ideal for analyzing extreme values and rare events. Each model addresses different data characteristics and analysis needs, providing versatile tools for statistical analysis. In practice, however, count data are often overdispersed to these distributions, and sometimes extreme observations are also present. For instance, the count data of upheld complaints of automobile insurance companies complaints rankings in New York City and the number of doctors visits to the hospital are heavy-tailed datasets and have excessive zeros. \answereditor{In both datasets, 47.53\% and 42\% of the values are zeros, indicating no upheld complaints registered and no recorded doctor visits at the hospital, respectively. Conversely, 6.18\% and 2.48\% of the observations are extreme values in both datasets, respectively, identified using the right fence with $c=3$ of the Tukey fence method \citep{schwertman2004simple}.} We fitted ZINB to both datasets and constructed their quantile-quantile (Q-Q) plots. Figure~\ref{fig:existmodel} (a, b) shows that the model fits well to the lower and central parts of the data distribution but exhibits deficiencies at the upper tail.  Consequently, building a model for the count data sets with ZI and extreme observations becomes a significant problem and has not yet been properly studied.


  Modeling discrete exceedances is a viable approach to analyzing extreme tail behavior, offering valuable insights into rare events.
 The peak-over-threshold (POT) method, as introduced by \citet{pickands1975statistical}, commonly approximates the distribution of exceedances above a high threshold to Generalized Pareto Distribution (GPD). Several adaptations of the GPD for discrete data are available in the literature \citep{krishna2009discrete, buddana2014discrete, kozubowski2015discrete}. More recently, \citet{hitz_davis_samorodnitsky_2024} discussed the DGPD for modeling the tail behavior of integer-valued random variables. Consequently, when the threshold is sufficiently high, standard extreme value models, such as the POT method with DGPD approximation \citep{ranjbar2022modelling, hitz_davis_samorodnitsky_2024}, can be applied to model the exceedances. Detailed information is provided in Section \ref{MF}. Choosing an appropriate threshold $u$ is critical for effectively applying the POT approximation. Setting the threshold too low can introduce bias into the estimates, as the DGPD is justified only in an asymptotic sense. On the other hand, setting the threshold too high reduces the number of data points, increasing the estimation variance. In practice, selecting a suitable threshold in a continuous setting often involves parameters stability plots and mean residual life plots, as suggested by \citet[ch. 4]{coles2001introduction} and \citet{davison1990models}. A suitable threshold can also be identified where the mean of the exceedances (mean residual life) starts to exhibit a linear relationship with the threshold value \citep{daouia:2023}. In the discrete POT approach, defining a threshold at a high quantile remains necessary, which poses challenges due to the discrete nature of the data.

 Selecting an appropriate threshold is often challenging in practice, as stability plots and mean residual life plots may not always clearly indicate the best threshold. Furthermore, thresholds that are straightforward for practitioners to interpret might be too low to accurately estimate extremes. When applying the POT model, this can lead to modeling discrete exceedances above a threshold that is too low from the standpoint of extreme value theory, thereby invalidating the approximation of the DGPD. An alternative method for modeling discrete exceedances above a low threshold involves using a discrete version of one of the three extended generalized Pareto models introduced by \citet{papastathopoulos2013extended}.

 In this contribution to the literature, we introduce new versions of the DGPD to enhance the model proposed by \citet{ahmad2024extended}. The proposed models exhibit flexibility in three key areas: First, effectively model the entire range of non-negative integer-valued data, accommodating a wide variety of count data scenarios; second, correctly handle ZI integer-valued data, providing accurate and reliable results even when datasets contain a significant proportion of zero counts; and, third, model the exceedances above a low threshold without requiring a censoring in the likelihood, simplifying the modeling process in cases where setting an appropriate high threshold is challenging. These enhancements offer a comprehensive and flexible approach to modeling discrete data, particularly in situations where existing models fail, see Figure~\ref{fig:existmodel}.

The remainder of the paper is structured as follows. Section~\ref{MF} introduces the extended versions of the DGPD, \answer{specifically the discrete extended generalized Pareto distribution (DEGPD) and zero-inflated DEGPD (ZIDEGPD). Section~\ref{dist-reg} presents the distributional regression framework using model parameters as a function of the available covariates.} Section~\ref{simstudy} presents the results of our simulation study, while Section~\ref{realap} covers the application of models to the real data. Lastly, Section \ref{conl} summarizes the findings and discusses potential directions for future research.

\section{Modeling framework}\label{MF}
Exceedances, which refer to the amount of data exceeding a specific high threshold, are commonly modeled using the GPD. \answer{The GPD is defined by its cumulative distribution function (CDF) in Definition~\ref{def1}.
\begin{mydefinition}\label{def1}
The CDF of GPD is written as 
\begin{equation} \label{CDFGPD}
	\mathcal{F}(z;\beta,\xi) =
	\begin{cases} 
		1-\left(1+\xi  z/\beta\right)_{+}^{-1/\xi} & \xi\neq 0 \\
		1- \exp{(-{z}/{\beta})} &  \xi = 0
	\end{cases}
\end{equation}
where $(a)_{+}= \max(a,0)$, $\beta>0$, and $-\infty<\xi<+\infty$ represent the scale and shape parameters of the distribution, respectively.  
\end{mydefinition}
 As the threshold increases, the limiting distribution of observations converges to a GPD as explained 
 below.

For a random variable $X$ that takes values in the interval in $[0, x_\mathcal{F} )$ where $x_\mathcal{F} \in(0,\infty)\cup \{\infty\}$\answerR{, 
 suppose} that there exists a strictly positive sequence $a_u$ such that the distribution of 
  $a_u^{-1}  (X - u) | X \ge u$ weakly \answerRrR{converges} to a non-degenerate probability distribution on $[0, \infty\answerR{)}$ as \answerR{$u\rightarrow
  x_\mathcal{F}$}, then this distribution is the GPD \citep{balkema:de_haan:1974}.
Thus, for sufficiently large $u$, 
 $$\Pr (X - u > x | X \ge u) = \Pr (a_u^{-1} (X - u) >  a_u^{-1} x | X \ge u)\approx 1-	\mathcal{F}(x;a_u\beta,\xi).$$ 
}
The shape parameter $\xi$ of the GPD characterizes the tail behavior. If $\xi<0$, the upper tail is bounded. If $\xi=0$, the distribution approaches an exponential form with all moments finite. If $\xi>0$, the upper tail is unbounded, but higher moments ultimately become infinite. These characteristics describe "short-tailed," "light-tailed," and "heavy-tailed" behaviors, respectively, making the GPD versatile for modeling excesses.

However, applying the GPD to discrete data can be problematic, as noted by
\citet{hitz_davis_samorodnitsky_2024}.
They suggested that for an integer-valued random variable $Y$, the tail distribution can be better approximated by discretizing the CDF \eqref{CDFGPD} for large $u$.
\answer{
\begin{mydefinition}
Let $Y$ be an integer-valued random variable and $u$ be a sufficiently high threshold. The discrete distribution is defined as
\begin{equation} \label{DGPD}
	\Pr(Y-u=k|Y\ge u) = \mathcal{F}(k+1;\beta,\xi)- \mathcal{F}(k;\beta,\xi), \hspace{1cm} k \in \mathbb{N}_{0},
\end{equation}
with  $\beta>0$ and $\xi\ge 0$, \answerRrR{the $\mathcal{F}(.;\beta,\xi)$ is the CDF of GPD defined in \eqref{CDFGPD}.} 
The distribution given in \eqref{DGPD} is known as the DGPD and has been explored in the literature from different perspectives; see, for instance, \citet[and references therein]{hitz_davis_samorodnitsky_2024}. 
\end{mydefinition}
}
A notable drawback of the DGPD is its reliance on observations exceeding a certain threshold, which imposes an artificial dichotomy on the data (i.e., values are classified as either above or below the threshold). This threshold selection can be complex, particularly with datasets containing many tied values. In contrast, some continuous extreme value models aim to avoid the need of threshold selection by modeling the entire range of data \citep{frigessi2002dynamic,carreau2009hybrid,macdonald2011flexible,papastathopoulos2013extended,naveau2016modeling,stein:2021}

A noteworthy extension to GPD proposed by \citet{papastathopoulos2013extended} incorporates an additional shape parameter without affecting the tail behavior. Additional shape parameter stabilized the threshold estimation for the GPD parameter, allowing a lower threshold to be chosen. In addition to modeling the lower tail and the bulk of the distribution, \citet{naveau2016modeling} identified two conditions to ensure compliance with the EVT. The transition between the two tails on $[0,1]$ can take various forms. We follow the footsteps of \citet{naveau2016modeling} to develop a model for an entire range of discrete data. They use the idea of integral transformation to simulate GPD random draws, i.e. $\mathcal{F}_{\beta,\xi}^{-1}( U),$ where $U\sim\mathcal{U}(0,1)$ represents a uniformly distributed random variable on $(0,1)$ and $\mathcal{F}_{\beta,\xi}^{-1}$ denotes the inverse of the CDF \eqref{CDFGPD}. Based on this, we can construct a family of distributions for the random variables 
	\begin{equation}\label{eq:3}
		Z=\mathcal{F}_{\beta,\xi}^{-1}\left(\mathcal{G}^{-1}(U) \right),
	\end{equation}
	where $\mathcal{G}$ is a CDF on $[0,1]$ and $U\sim\mathcal{U}(0,1)$. Clearly the CDF of $Z$ is $\mathcal{G}(\mathcal{F}(z;\beta,\xi))$. The key problem is to find a function $\mathcal{G}$, which preserves the upper tail behavior with shape parameter $\xi$ and also controls the lower tail behavior. \citet{naveau2016modeling} defined   restrictions for validity of $\mathcal{G}$ functions. For instance, the tail of $\mathcal{G}$  denoted  by $\Bar{\mathcal{G}}=1-\mathcal{G}$ has to satisfy 
\begin{eqnarray}
	\lim_{\nu \to 0} \frac{\Bar{\mathcal{G}}(1-\nu)}{\nu}&=&a,  \mbox{ for some finite $a>0$ (upper tail behavior),}\label{eq:cond1}\\
	\lim_{\nu \to 0} \frac{\mathcal{G}(\nu)}{\nu^\kappa}&=&c,  \mbox{ for some finite $c>0$ (lower tail behavior).}\label{eq:cond2}
\end{eqnarray}
\answer{
\begin{mydefinition}\label{degpd-def}
    Let $Y$ be an integer-valued random variable. The probability mass function (PDF) of DEGPD is defined by using definition~\ref{def1} as
\begin{equation}\label{eq:DEGPD}
	\Pr(Y=k) = \mathcal{G}\left(\mathcal{F}\left({k+1};\beta,\xi\right);\kappa  \right)- \mathcal{G}\left(\mathcal{F}\left({k};\beta,\xi\right);\kappa  \right),\qquad  k= 0, 1, 2,\ldots,
\end{equation}
\answerRrR{where the parameter $\kappa$ represent the lower tail of the distribution which modeled by $\mathcal{G}$ defined in Models (i)-(iii) later, $\beta$ and $\xi$ are usual parameters of GPD. The $\mathcal{G}\left(\mathcal{F}\left({.};\beta,\xi\right);\kappa  \right)$ is the CDF of EGPD in accordance with Models (i)-(iii).
For $\xi>0$, we have shown that the EGPD CDF satisfies extreme value theory properties, exhibiting a regularly varying tail. Moreover, the model introduced in~\eqref{eq:DEGPD} is also regularly varying, sharing the same tail behavior as in the continuous case, and belongs to the maximum domain of attraction. Detailed proofs of the EGPD and DEGPD properties for each Model (i)–(iii) are provided in the Supplementary Material.} The explicit formula of the CDF of DEGPD is

\begin{equation}\label{CDFDEGPD}
 	\Pr(Y\le k)= \mathcal{G}\left(\mathcal{F}\left({k+1};\beta,\xi\right), \kappa \right),
 \end{equation} 
 and quatntile function is \begin{equation}\label{QDEGPD}
 	q_{p}=
 \left\{
 	\begin{array}{ll}
 		\left\lceil{\frac{\beta}{\xi}\left\{ \left(1- \mathcal{G}^{-1}(p) \right)^{-\xi}-1\right\}}\right\rceil-1, & \mbox{if } \xi>0\\
   \\
 		\left\lceil{-\beta \log \left(1- \mathcal{G}^{-1}(p) \right)}\right\rceil -1,& \mbox{if } \xi=0.\\
 	\end{array}
 \right.
 \end{equation} 
\end{mydefinition}
}
We define another PDF by following \citet{lambert1992zero} for the case of zero-inflation in count data.
\answer{
\begin{mydefinition}
Suppose $Y$ is observed with excessive zeros relative to those observed under the \eqref{eq:DEGPD}. In that case, the PDF with additional $0\leq \pi \leq 1$ parameter, which represent zero values, is defined as 
\begin{equation}\label{eq:ZIDEGPD}
	\Pr(Y=y) = \left\{
		\begin{array}{ll}
			\pi +(1-\pi)\mathcal{G}\left(\mathcal{F}\left(1;\beta,\xi\right); \kappa \right) & k=0 \\
			(1-\pi)\left[\mathcal{G}\left(
			\mathcal{F}\left(k+1;\beta,\xi\right); \kappa \right) -\mathcal{G}\left(\mathcal{F}\left(k;\beta,\xi\right); \kappa \right)
			\right] & k= 1,2,\ldots
		\end{array}
		\right.
\end{equation}
The distribution defined by \eqref{eq:ZIDEGPD} is referred to as the ZIDEGPD. The explicit formula of the CDF of ZIDEGPD is written as

\begin{equation}\label{eq:11}
		\Pr(Y\leq k)=
		\pi +(1-\pi)\mathcal{G}\left(\mathcal{F}\left(k+1,\beta,\xi\right),\kappa \right),\qquad  k=0, 1, 2,\ldots, 
	\end{equation}
and the quantile function is derived
 \begin{equation}\label{QZIDEGPD}
		q_{p^{*}}=\left\{
		\begin{array}{lc}
			\ceil{\frac{\beta}{\xi}\left[ \left\{1- \mathcal{G}^{-1}(p^{*}) \right\}^{-\xi}-1\right]}-1, &  \xi>0\\
			\lceil{-\beta \log \left\{1- \mathcal{G}^{-1}(p^{*}) \right\}}\rceil-1, 
			& \xi=0,\\
		\end{array}
		\right.
	\end{equation} 
	with $0<p<1$ and $0<p^{*}=(p-\pi)/(1-\pi)<1$.
\end{mydefinition}
}

 In this paper,  we use  three parametric expressions for $\mathcal{G}(\cdot)$ that fulfill the conditions \eqref{eq:cond1} and conditions $(C_1, C_2$ and $C_3)$ presented in \citet{gamet2022}. The following three models fulfill the above-mentioned conditions and are deemed suitable for modeling the bulk of the distribution while maintaining the stability of heavier tail behavior. 
\begin{description}
	\item[Model (i) $(M_1)$: ] \label{c1} 
	$\mathcal{G}(\nu; \kappa)=\nu^{\kappa}$, $\kappa >0$;
	\item[Model (ii) $(M_2)$:]\label{c3}
	$\mathcal{G}(\nu; \kappa)=\frac{2}{2\Phi\left(\sqrt{\kappa}\right)-1} \Big[\Phi\left(\sqrt{\kappa}(\nu-1)\right)-\left\{1-\Phi(\sqrt{\kappa})\right\}\Big]$, the defined $\mathcal{G}(\nu)$ is the CDF of truncated normal distribution with precision $\kappa>0$ and $\Phi$ is the CDF of standard normal distribution. After satisfying the conditions $(C_1, C_2$ and $C_3)$, the density of function of \answerR{truncated normal distribution} for variable $U$ is defined as
 $$f_{\nu}(\nu; \kappa)= \left\{
		\begin{array}{ll}
			\frac{2 \sqrt{\kappa}}{2\Phi\left(\sqrt{\kappa}\right)-1}\phi\left(\sqrt{\kappa}(\nu-1)\right)  & \text{if}\hspace{0.5cm} \kappa>0 \\
			1 & \text{if} \hspace{0.5cm}\kappa \rightarrow 0,
		\end{array}
		\right.$$
  \answerRRR{where $\phi$ is the density function of the standard normal
distribution.} The rest of the article will refer to the model associated with this $ \mathcal{G}(.)$ choice as DEGPD-Normal and ZIDEGPD-Normal.
	\item[Model (iii) $(M_3)$:] \label{c4}
 We consider the truncated Beta distribution as an alternative choice to add more flexibility to the density of $U$. Let $D_{x}(\alpha_1, \alpha_2)$ indicates incomplete beta function:
	$$D_{x}(\alpha_1, \alpha_2)=\int_{0}^{x}t^{\alpha_1 -1}(1-t)^{\alpha_2 -1} dt,$$ 
 and the density function of the Beta distribution is written as 
 $$\text{Beta}(\nu; \alpha_1, \alpha_2)= \frac{\Gamma(\alpha_1+ \alpha_2)}{\Gamma(\alpha_1)\Gamma (\alpha_2)} \nu^{\alpha_1-1}(1-\nu)^{\alpha_2-1},$$
 with $0\leq \nu \leq 1, \alpha_1>0$ and $\alpha_2>0$.The distribution with parameters $(\kappa, \kappa)$ bounded on interval $(\omega, 1/2)$ with $\kappa>0$ and $0<\omega<1/2$ 
 \begin{equation}\label{betaden}
     f_{\nu}(\nu; \kappa)= \frac{1}{D_{1/2}(\kappa, \kappa)-D_{\omega}(\kappa, \kappa)} Beta\left((1/2-\omega)\nu+\omega; \kappa, \kappa\right).
 \end{equation}
 The expression \eqref{betaden} rescaled to the unit interval and satisfied the conditions $(C_1, C_2$ and $C_3)$. For more details, we refer to see, for instance, \citet {gamet2022}. The CDF corresponding to the PDF defined in \eqref{betaden} is written as 
 \begin{equation}
     \mathcal{G}(\nu;\kappa)= \left\{
		\begin{array}{ll}
			0  & \text{for}\hspace{0.5cm} \nu<0; \\
			\frac{D_{\left\{(1/2 -\omega)\nu-\omega\right\}}(\kappa, \kappa)- D_{\omega}(\kappa, \kappa)}{D_{1/2}(\kappa, \kappa)-D_{\omega}(\kappa, \kappa)} & \text{for} \hspace{0.5cm} 0 \leq \nu \leq 1;\\
   1  & \text{for}\hspace{0.5cm} \nu>1.
		\end{array}
  \right.
 \end{equation}
\end{description}
\begin{figure}
    \centering
\includegraphics[width=0.48\linewidth]{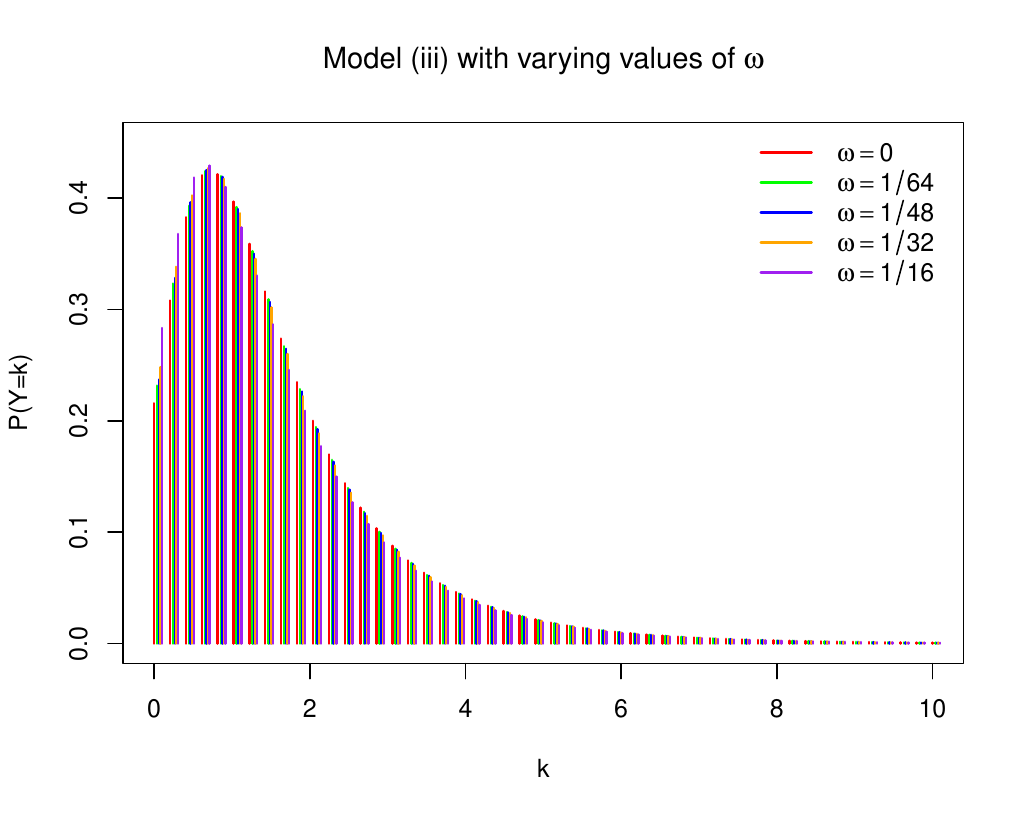}
    \includegraphics[width=0.48\linewidth]{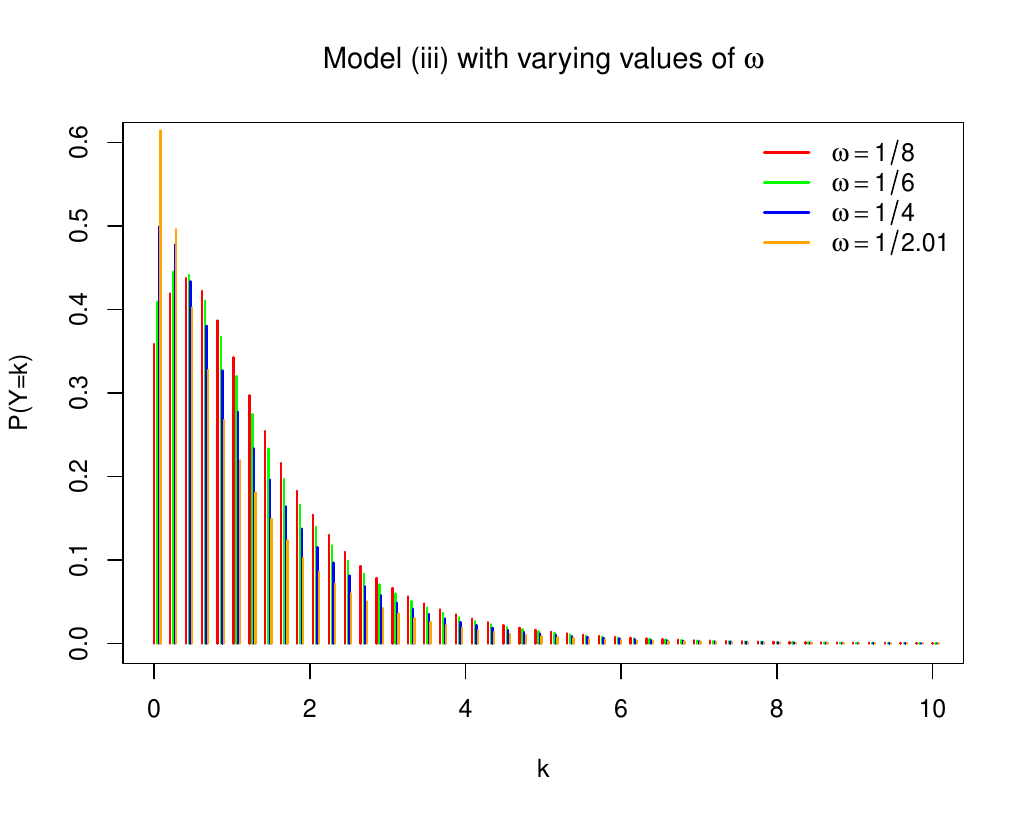}
    \caption{Flexibility check for model $M_3$ with varying $\omega$ values and keeping the values of other parameters as $\kappa=5, \beta=1,$ and $\xi=0.1$. In left panel $\omega\in(0, 1/16)$ while in right panel $\omega\in(1/8, 1/2.01)$, where $2.01$ ensure the choice of $\omega$ is not exactly 0.5.  }
    \label{fig:enter-label}
\end{figure}
\answerRRR{To choose the appropriate value of $\omega$, we have check the flexibility of the model $M_3$ in Figure \ref{fig:enter-label} with different values of $\omega$. In the left panel, it can be seen that the model is flexible when varying the values of $\omega$ from $(0,1/64, 1/48, 1/32, 1/16)$. The behavior is almost similar in most of the distribution in the bulk part, and the tail is constant. On the other hand, when varying $\omega$ $(1/8, 1/2)$, the model loses flexibility in the bulk part of the distribution.  It turns out that $\omega=1/32$ leads to a flexible model for low values with no change in the right tail. If the model is not flexible enough in some real applications, the value could be set to another value, as shown in the right panel of Figure \ref{fig:enter-label}, or could be estimated from the data.}
Nevertheless, this may cause identifiability problems in the later scenario. This model will be referred to as DEGPD-Beta and ZIDEGPD-Beta in the paper.

\begin{figure}
\centering
\subfloat[DEGPD-Power]{%
\resizebox*{4.6cm}{!}{\includegraphics{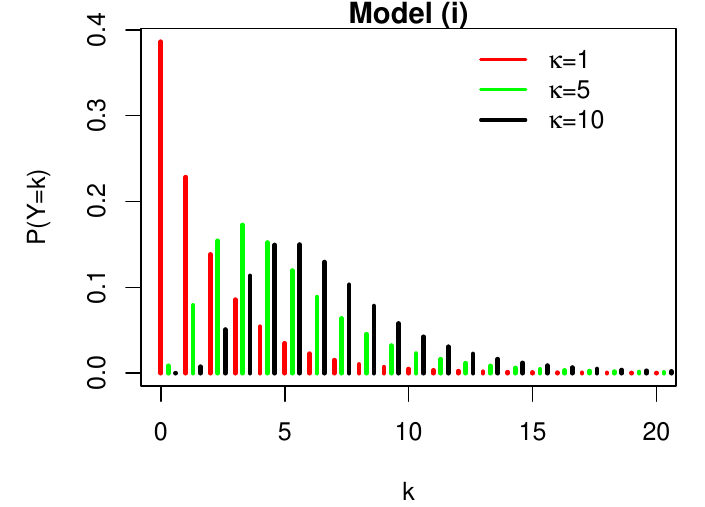}}}\hspace{5pt}
\subfloat[DEGPD-Normal]{%
\resizebox*{4.6cm}{!}{\includegraphics{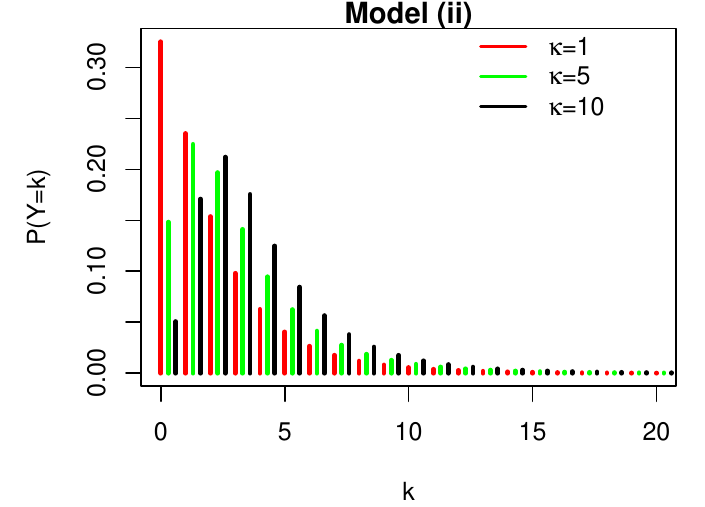}}}
\subfloat[DEGPD-Beta]{%
\resizebox*{4.6cm}{!}{\includegraphics{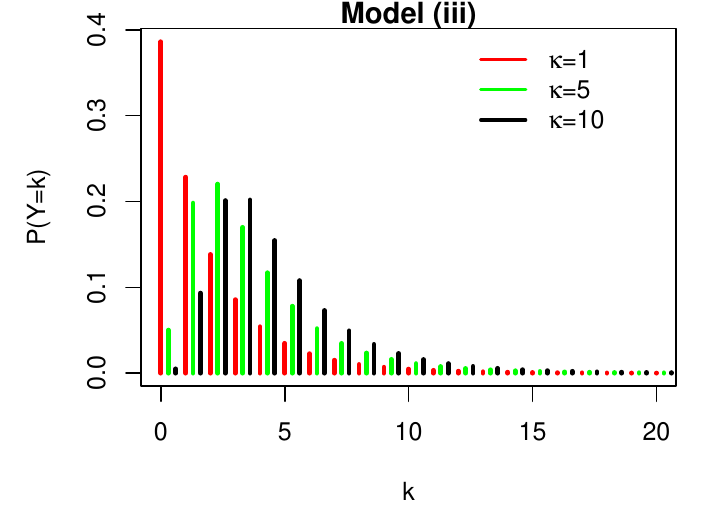}}}\\
\subfloat[ZIDEGPD-Power]{%
\resizebox*{4.6cm}{!}{\includegraphics{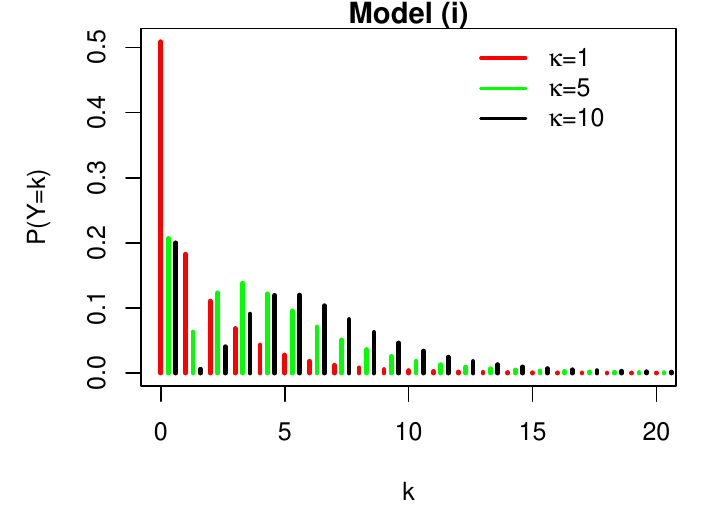}}}\hspace{5pt}
\subfloat[ZIDEGPD-Normal]{%
\resizebox*{4.6cm}{!}{\includegraphics{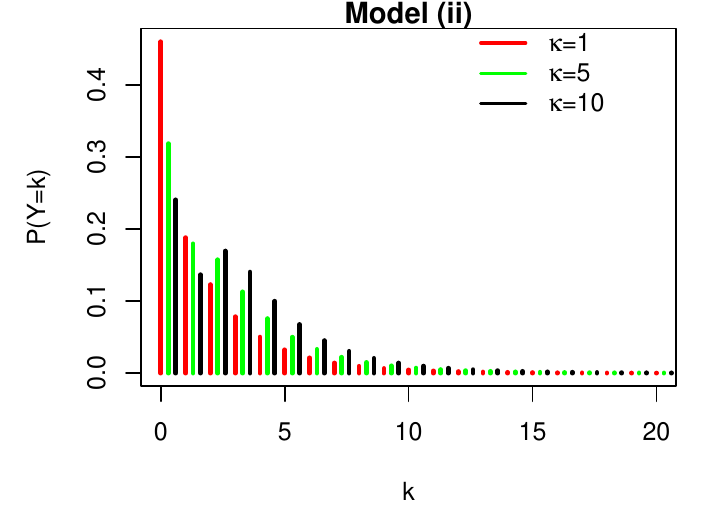}}}
\subfloat[ZIDEGPD-Beta]{%
\resizebox*{4.6cm}{!}{\includegraphics{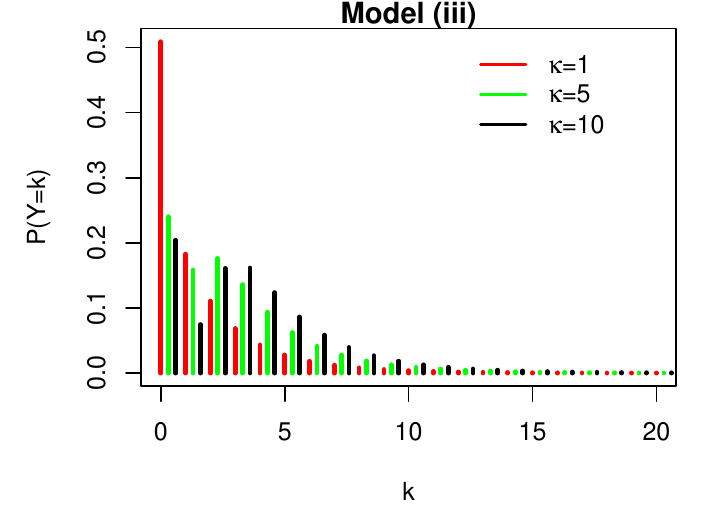}}}
\caption{Probability mass function behavior for DEGPD and ZIDEGPD based on $M_1, M_2$ and $M_3$.}
	\label{fig:DENSBEHHAV}
\end{figure}

The models $M_1, M_2$ and $M_3$ by mixture of \eqref{eq:DEGPD} and \eqref{eq:ZIDEGPD} lead to a PDF of DEGPD with parameters $(\kappa, \beta$ and $\xi)$ and ZIDEGPD with parameters$(\pi, \kappa, \beta$ and $\xi)$: $\kappa$ deals the shape of the lower tail, $\beta$ is a scale parameter, and $\xi$ controls the rate of
upper tail decay. The additional parameter $\pi$ in ZIDEGPD represents the proportion of excessive zeros in the data. Figure~\ref{fig:DENSBEHHAV} (top) shows the behavior of PDF of DEGPD for $M_1, M_2$ and $M_3$ with fixed scale and upper tail shape parameter (i.e., $\beta=1$ and $\xi=0.2$) and with different values of lower tail behaviors ($\kappa=1, 5$ and $10$). Further, Figure~\ref{fig:DENSBEHHAV} (bottom) illustrates the PDF of ZIDEGPD for $M_1, M_2$ and $M_3$ with fixed $\pi$, scale and upper tail shape parameter (i.e., $\pi=0.2$ $\beta=1$ and $\xi=0.2$) and with different values of lower tail behaviors ($\kappa=1, 5$ and $10$). 
The DGPD and ZIDGPD can be recovered when $\kappa=1$ in model $M_1$ and $M_3$ and $\kappa\rightarrow 0$ in $M_2$. When varying $\kappa$, the additional flexibility for lower tail can be observed without losing upper tail behavior. In the case of the zero-inflation model, $M_1, M_2$ and $M_3$ yield zeros even when $\kappa=10$.



To estimate the parameters of the models proposed in \eqref{eq:DEGPD} and \eqref{eq:ZIDEGPD}, the maximum likelihood estimation (MLE) method is practiced. 
Let $y_1, \ldots, y_n$ be $n$ independent observations from \eqref{eq:DEGPD}. The log-likelihood function is given by, 
\begin{equation}\label{eq:lik}
	l(\boldsymbol{\theta})= \sum_{i=1}^{n} \log \left[ \mathcal{G}(\mathcal{F}(y_{i}+1; \beta, \xi);\kappa)- 
	\mathcal{G}(\mathcal{F}(y_{i}; \beta, \xi);\kappa) \right].
\end{equation}


Instead if we consider $n$ independent observations $y_1,\ldots,y_n$ from \eqref{eq:ZIDEGPD}
we get
\begin{eqnarray}\label{eq:likzi}
	l(\boldsymbol{\theta})&=& \sum_{i=1}^n I_0(y_i)\log \left[\pi+(1-\pi)
				\mathcal{G}(\mathcal{F}(1;\beta, \xi);\kappa)\right] \nonumber \\ 
	&&+\sum_{i=1}^{n}(1-I_0(y_i)) \log(1-\pi))\times 
	 \left[ \mathcal{G}(\mathcal{F}(y_{i}+1;\beta, \xi);\kappa)- 
	\mathcal{G}(\mathcal{F}(y_{i};\beta, \xi);\kappa) \right],
\end{eqnarray}
where $\boldsymbol{\theta}$ is a vector of unknown parameters.  Derivatives concerning unknown parameters of DEPGD and ZIDEGPD can be solved by standard numerical techniques to obtain the maximum likelihood
estimators for unknown parameters.

\answerRR{\section{Distributional regression framework}\label{dist-reg}
In distributional regression setting, we consider the parameters of distribution $Y$ vary according to the covariate $\boldsymbol{W}=(w_1,\cdots, w_p)^\top$,
modeled as \( Y |\boldsymbol{w} \ \sim \text{DEGPD}(\cdot, \theta(\boldsymbol{w})) \) or \( Y |\boldsymbol{w} \ \sim \text{ZIDEGPD}(\cdot, \theta(\boldsymbol{w})) \), where \( \theta(\boldsymbol{w}) = (\theta_1(\boldsymbol{w}), \dots, \theta_d(\boldsymbol{w}))^\top \). This framework represents a distributional regression model \citep{stasinopoulos2018gamlss, carrer2022distributional}, where each parameter \( \theta_i(\boldsymbol{w}), i=1, \cdots, d \) is linked to covariates via an additive predictor:
\begin{equation}\label{eta}
\eta_i(w) = g_{i1}(\boldsymbol{w}) + \dots + g_{iJ_i}(\boldsymbol{w}),
\end{equation}
with smooth functions \( g_{ij}(\cdot), j=1,\cdots J_i \) capturing complex dependencies. For instance, with covariates \( (w_1, w_2) \), we might have \( \eta_i(\boldsymbol{w}) = \beta_{i1} + \beta_{i2} w_1 + g_{i2}(w_2) \), or for space-time data at location \( (s_1, s_2) \) and time \( t \), \(\boldsymbol{x} = (s_1, s_2, t)^\top \), we can set \( \eta_i(\boldsymbol{x}) = g_{i1}(t) + g_{i2}(s_1, s_2) \).
The link functions \( h_i(\cdot) \) relate \( \eta_i(\boldsymbol{x}) \) to distribution parameters. For example, the link functions considered for the DEGPD parameters are $\kappa(\boldsymbol{w})= \exp{( \eta_{\kappa}(\boldsymbol{w})})$, $\sigma(\boldsymbol{w})= \exp{( \eta_{\sigma}(\boldsymbol{w})})$ and $\xi(\boldsymbol{w})=  \eta_{\xi}(\boldsymbol{w})$ (i.e., identity link). In the case of ZIDEGPD, we additionally use a logit link for the parameter $\pi(\boldsymbol{w})$ as $\pi(\boldsymbol{w})=\exp{( \eta_{\pi}(\boldsymbol{w})})/1+\exp{( \eta_{\pi}(\boldsymbol{w})})$.
The functions \( g_{ij}(\boldsymbol{x}) \) in \eqref{eta} are approximated using basis expansions,
\[
g_{ij}(x) = \sum_{k=1}^{K_{ij}} \beta_{ij,k} B_k(x),
\]
with smoothing penalties \( \lambda_{ij} \beta_{ij}^\top W_{ij} \beta_{ij} \) added to the log-likelihood given in \eqref{eq:lik} or \eqref{eq:likzi}, yielding the penalized objective function
\begin{equation}\label{pen-link}
l_p = l(\boldsymbol{\theta}) - \frac{1}{2} \sum_{i=1}^{d} \sum_{j=1}^{J_i} \beta_{ij}^\top W_{ij} \beta_{ij}.   
\end{equation}
One can estimate the equation \eqref{pen-link} for model $M_2$ and $M_3$ by extending the code provided for $M_1$ in \citet{ahmad2024extended}. The functions developed for fitting $M_1$ are easy to use for $M_2$ and $M_3$ that facilitate the use of thin plate regression splines via \texttt{evgam} package, which is highly advantageous for modeling complex multidimensional processes. These splines are particularly suitable for spatial modeling and the representation of interaction effects by constructing tensor products. Due to space and data limitations, this paper has not implemented the distributional regression framework, but interested readers can easily apply it. }

\section{Simulation study}\label{simstudy}
\subsection{Model based simulation study}
This section presents a simulation study evaluating the performance of the maximum likelihood estimator (MLE). Different parameter settings were tested to determine the MLE for the DEGPD and ZIDEGPD models. Moreover, the scale parameter $\beta$ is permanently set to one for all models. The other parameters, $\pi, \kappa$ and $\xi$, vary with the scenarios. A sample size of $n=1000$ with $10^4$ replications was used to estimate the parameters; this sample size choice was more representative of real-world applications. \answerRrR{During the analysis, the initial values for the DEGPD parameters $c(\kappa, \beta, \xi)$ were set as
\(
c\Big(\kappa = \sqrt[4]{\mathrm{mean}(Y)},\; \beta = \sqrt[4]{\mathrm{sd}(Y)},\; \xi = 0.2 \Big),
\)
and for the ZIDEGPD parameters $c(\pi, \kappa, \beta, \xi)$ as
\(
c\Big(\pi = 0.2,\; \kappa = \sqrt[4]{\mathrm{mean}(Y)},\; \beta = \sqrt[4]{\mathrm{sd}(Y)},\; \xi = 0.2 \Big),
\)
where $Y$ denotes the data generated from the model under consideration.}

\begin{figure}
\centering
\subfloat[DEGPD-Power]{%
\resizebox*{4.5cm}{!}{\includegraphics{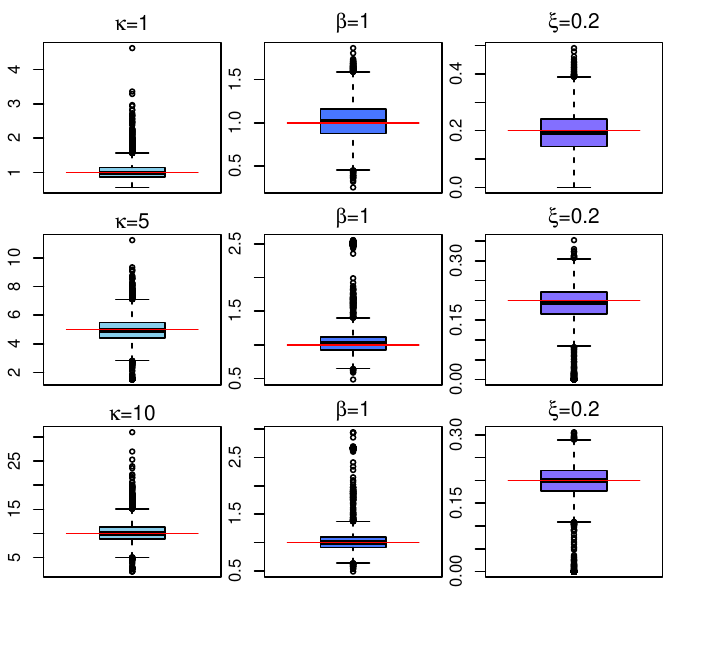}}}\hspace{5pt}
\subfloat[DEGPD-Normal]{%
\resizebox*{4.5cm}{!}{\includegraphics{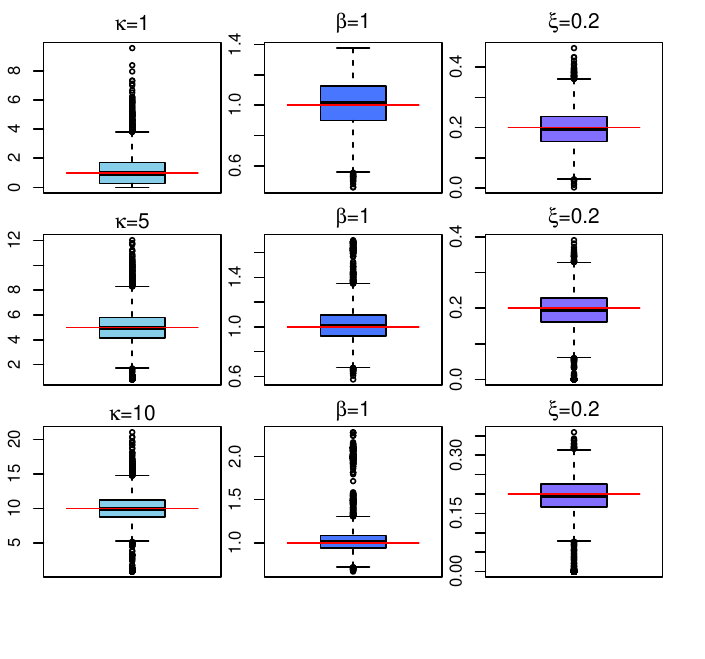}}}
\subfloat[DEGPD-Beta]{%
\resizebox*{4.5cm}{!}{\includegraphics{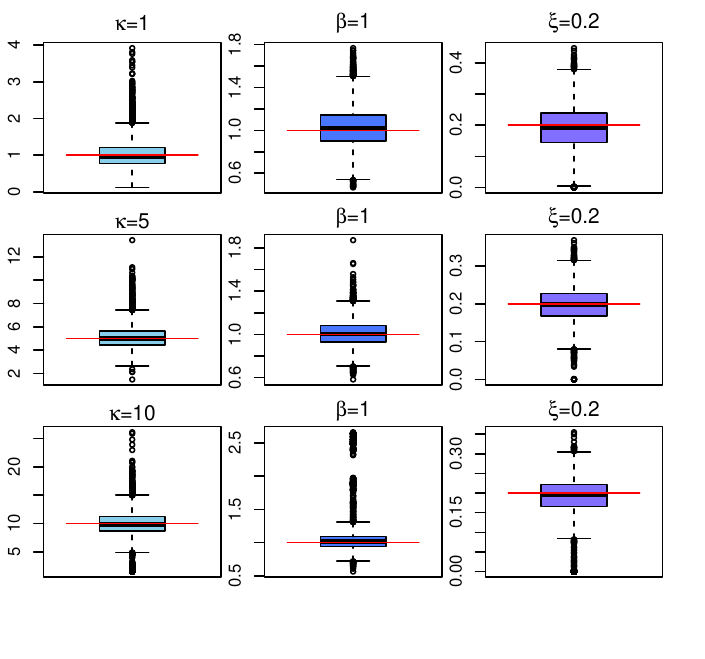}}}\\
\subfloat[ZIDEGPD-Power]{%
\resizebox*{4.5cm}{!}{\includegraphics{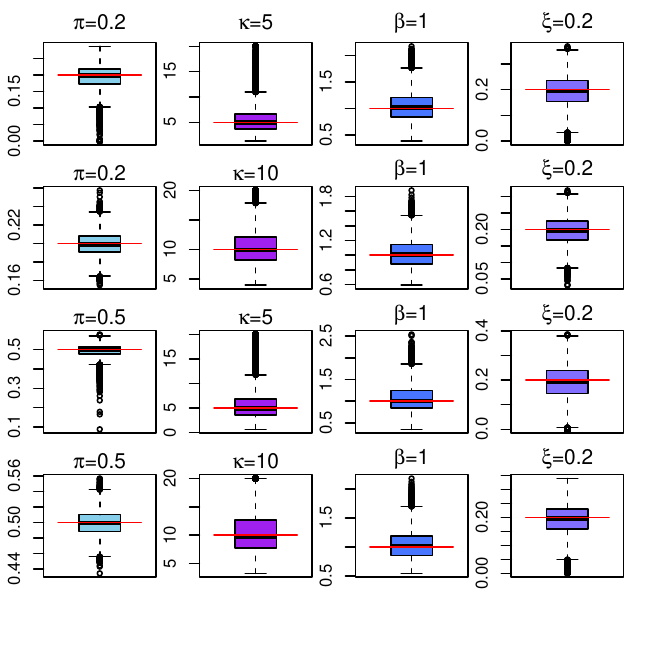}}}\hspace{5pt}
\subfloat[ZIDEGPD-Normal]{%
\resizebox*{4.5cm}{!}{\includegraphics{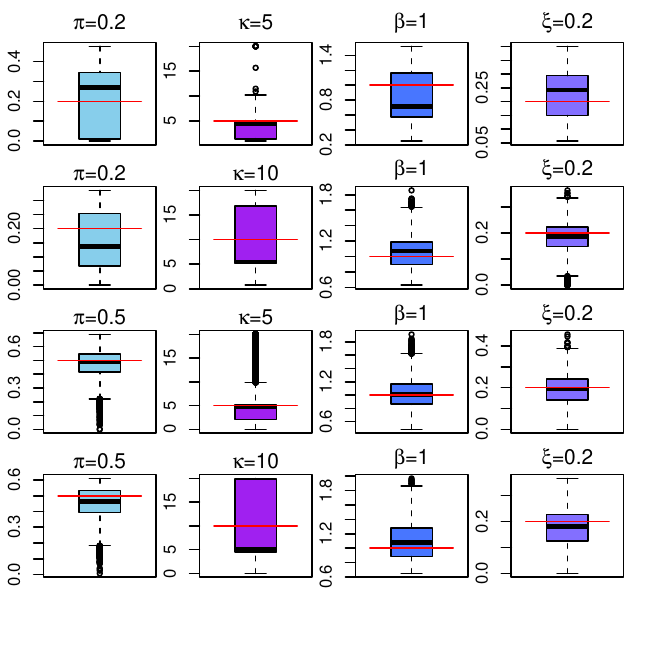}}}
\subfloat[ZIDEGPD-Beta]{%
\resizebox*{4.5cm}{!}{\includegraphics{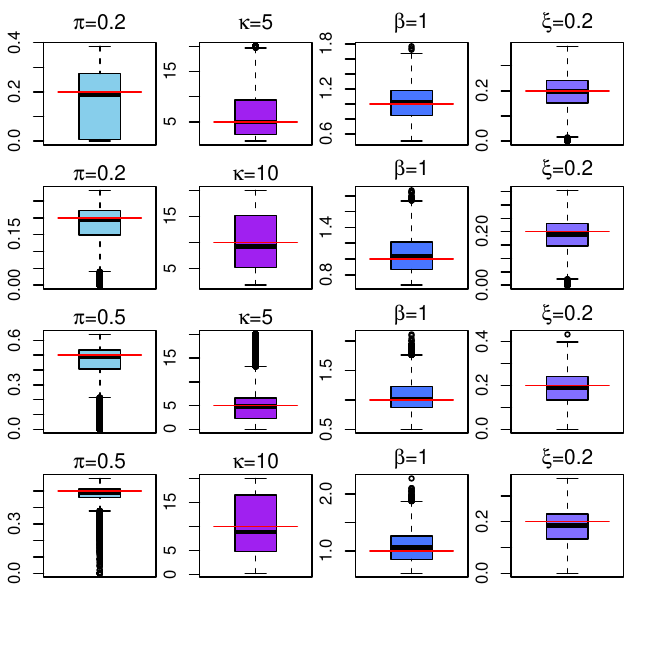}}}
\caption{Boxplots of estimated parameters using ML, (top) for DEGPD models $M_1$, $M_2$ and $M_3$ with parameters $\beta=1$, $\xi=0.2$ and $\kappa=1, 2, 10$. (Bottom) for ZIDEGPD models $M_1$, $M_2$ and $M_3$ with parameters $\pi=0.2, 0.5, \beta=1, \xi=0.2$ and $\kappa=5, 10$. Boxplots are based on $n=1000$ sample points with $10^4$ independent replicates, and horizontal red lines represent true values.}
	\label{fig:Simstudy}
\end{figure}

The boxplots of the  MLEs of the parameters are constructed to assess the performance for $M_1, M_2$, and $M_3$. Figure \ref{fig:Simstudy} (top) shows boxplots of the estimated parameters conforming to DEGPD and (bottom) to ZIDEGPD, along with all three models with different simulation settings. In each boxplot, the horizontal red line indicates the true value of the parameter. More precisely, the upper panel of  Figure \ref{fig:Simstudy} (left to right) indicates that the  MLEs based on $M_1, M_2$ and $M_3$ are reasonable and very close to the true parameters and do not show high variability even when the value of $\kappa$ is higher.

\begin{figure}
\centering
\subfloat[DEGPD-Power]{%
\resizebox*{4.5cm}{!}{\includegraphics{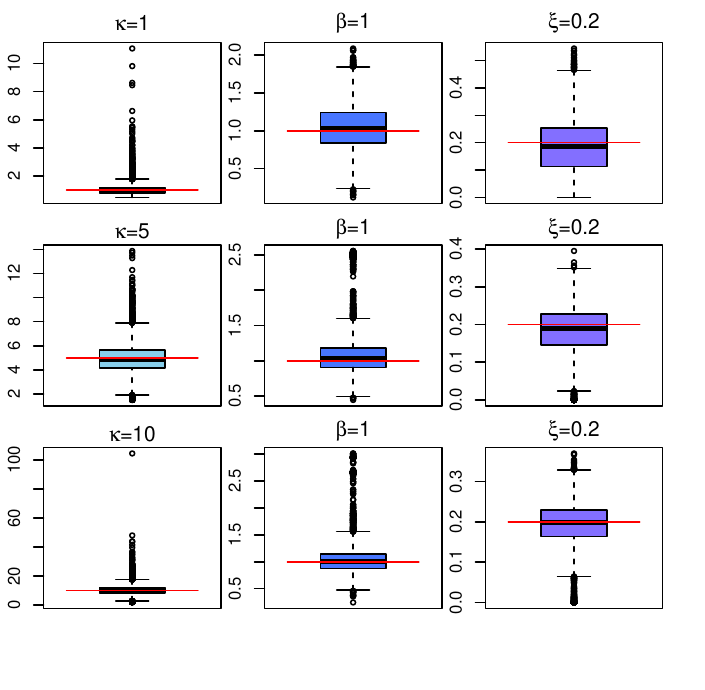}}}\hspace{5pt}
\subfloat[DEGPD-Normal]{%
\resizebox*{4.5cm}{!}{\includegraphics{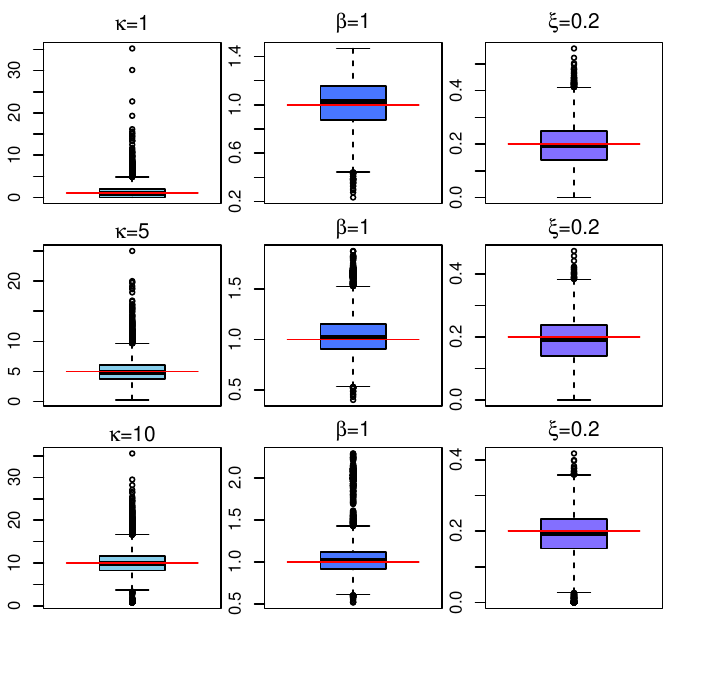}}}
\subfloat[DEGPD-Beta]{%
\resizebox*{4.5cm}{!}{\includegraphics{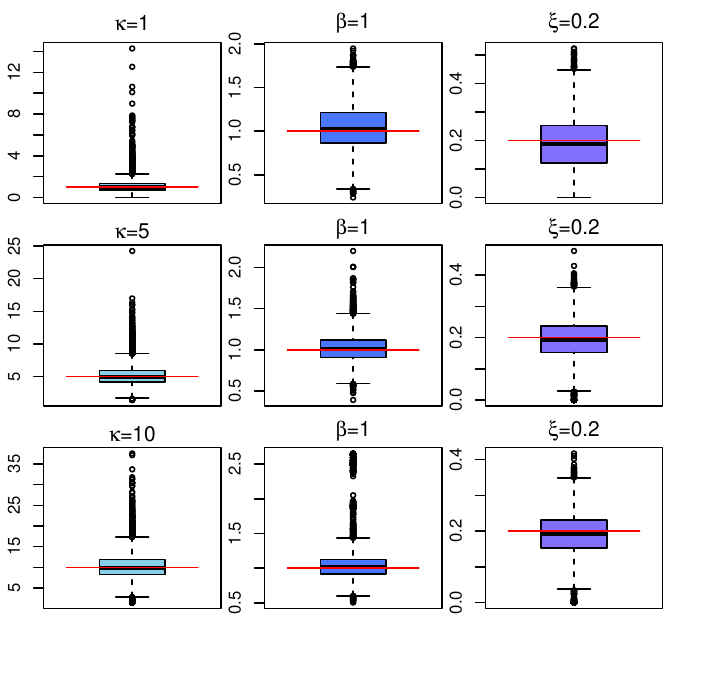}}}\\
\subfloat[ZIDEGPD-Power]{%
\resizebox*{4.5cm}{!}{\includegraphics{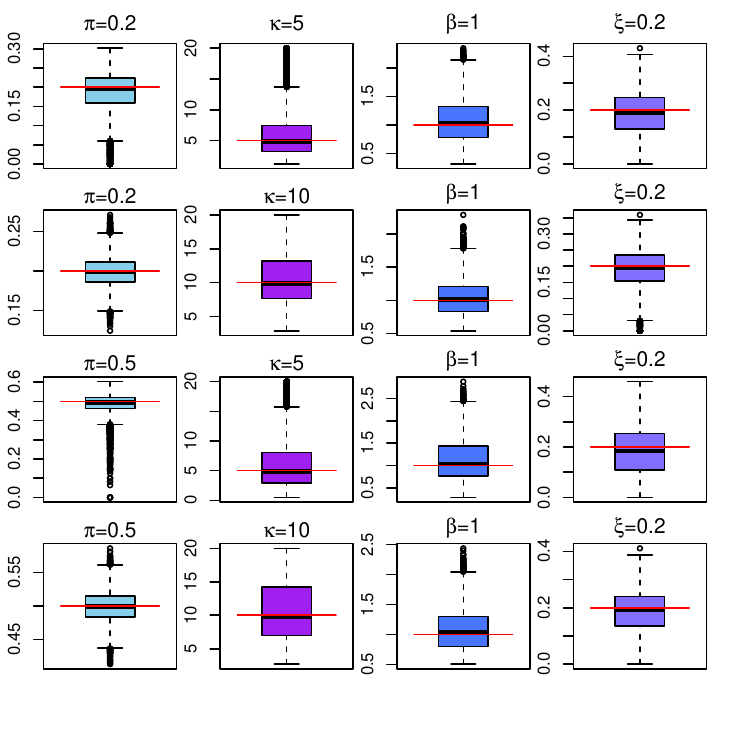}}}\hspace{5pt}
\subfloat[ZIDEGPD-Normal]{%
\resizebox*{4.5cm}{!}{\includegraphics{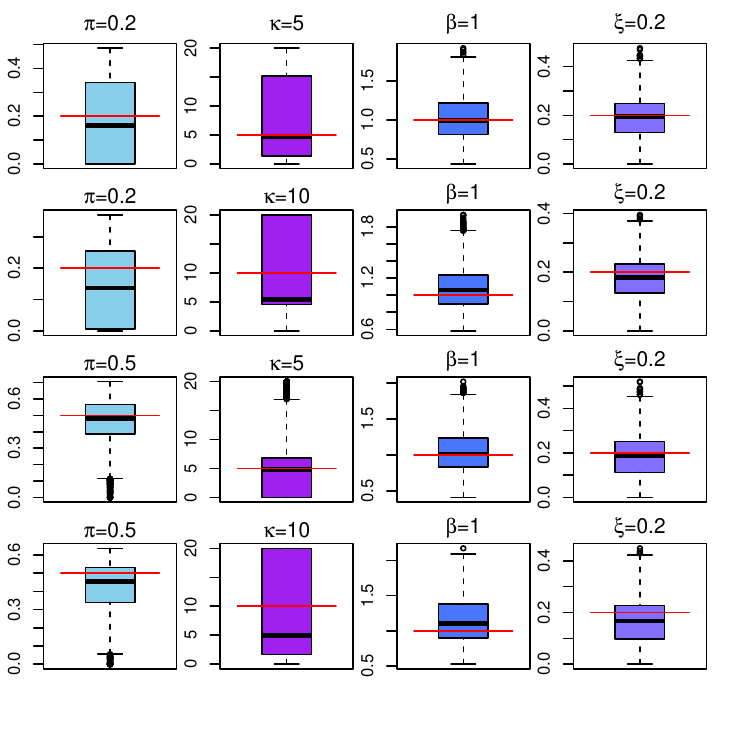}}}
\subfloat[ZIDEGPD-Beta]{%
\resizebox*{4.5cm}{!}{\includegraphics{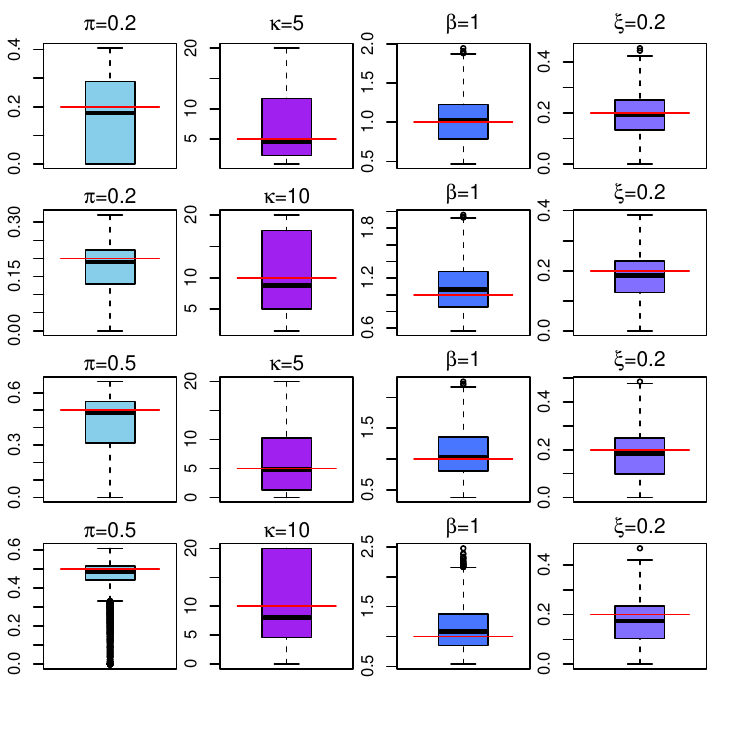}}}
\caption{Boxplots of estimated parameters using ML, (top) for DEGPD models $M_1$, $M_2$, and $M_3$ with parameters $\beta=1$, $\xi=0.2$, and $\kappa=1, 2, 10$. (Bottom) for ZIDEGPD models $M_1$, $M_2$ and $M_3$ with parameters $\pi=0.2, 0.5, \beta=1, \xi=0.2$ and $\kappa=5, 10$. Boxplots are based on $n=500$ sample points with $10^4$ independent replicates, and horizontal red lines represent true values.}
	\label{fig:Simstudy-n=500}
\end{figure}

In the bottom panel of Figure \ref{fig:Simstudy} (left to right), the estimated parameters of ZIDEGPD-Normal deviate from the true values when using ($\pi=0.2$ and $\kappa=5$). This result was expected because the simulation did not produce many zeros with $\pi=0.2$ choice. Secondly, this may happen due to truncation at zero and the combined influence of both  $\pi$ and $\kappa$ on the lower tail. However, for the other ZIDEGPD models based on $M_1$ and $M_3$, the estimated parameters appear fairly accurate. They remain close to the actual values and do not exhibit significant variability, even when $\pi$ is set to 0.5. This indicates that the MLE method performs well in estimating parameters for these scenarios, highlighting its robustness in dealing with various combinations of parameters in the ZIDEGPD models. \answerRR{The performance of MLEs was also tested with a sample size of $n=500$, and the results, shown in Figure~\ref{fig:Simstudy-n=500}, exhibit a similar pattern to those observed for $n=1000$.}

Overall, the simulation study provides valuable insights into the performance of the MLE in estimating the parameters for the DEGPD and ZIDEGPD models under different conditions. It helps practitioners understand the strengths and limitations of this estimation technique, particularly in scenarios where zero inflation and other distributional complexities come into play.

\begin{figure}
\centering
\subfloat[$\kappa$]{%
\resizebox*{6cm}{!}{\includegraphics{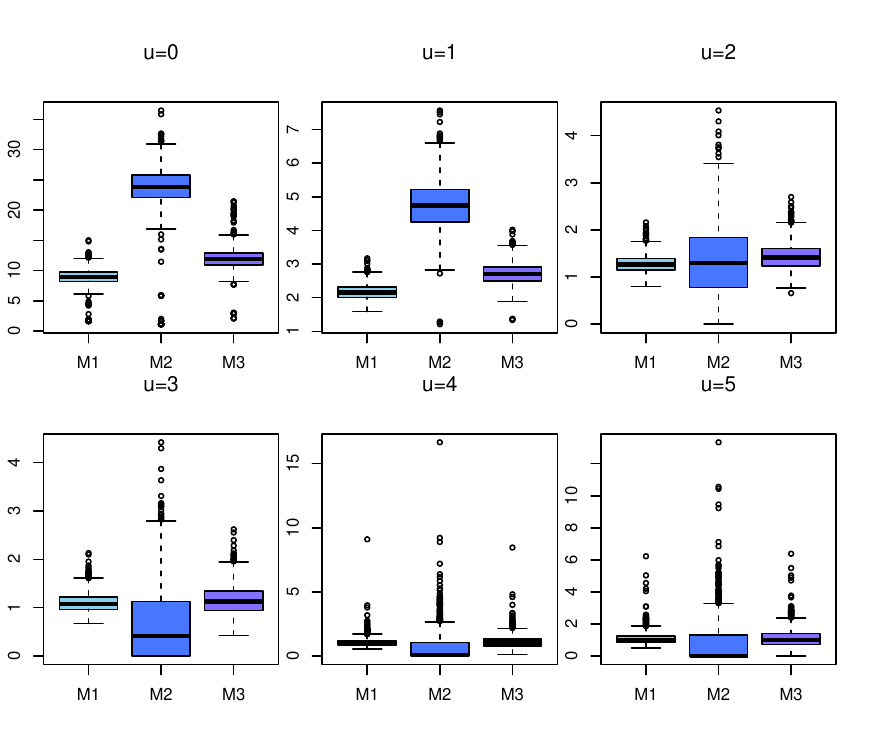}}}\hspace{5pt}
\subfloat[$\beta$]{%
\resizebox*{6cm}{!}{\includegraphics{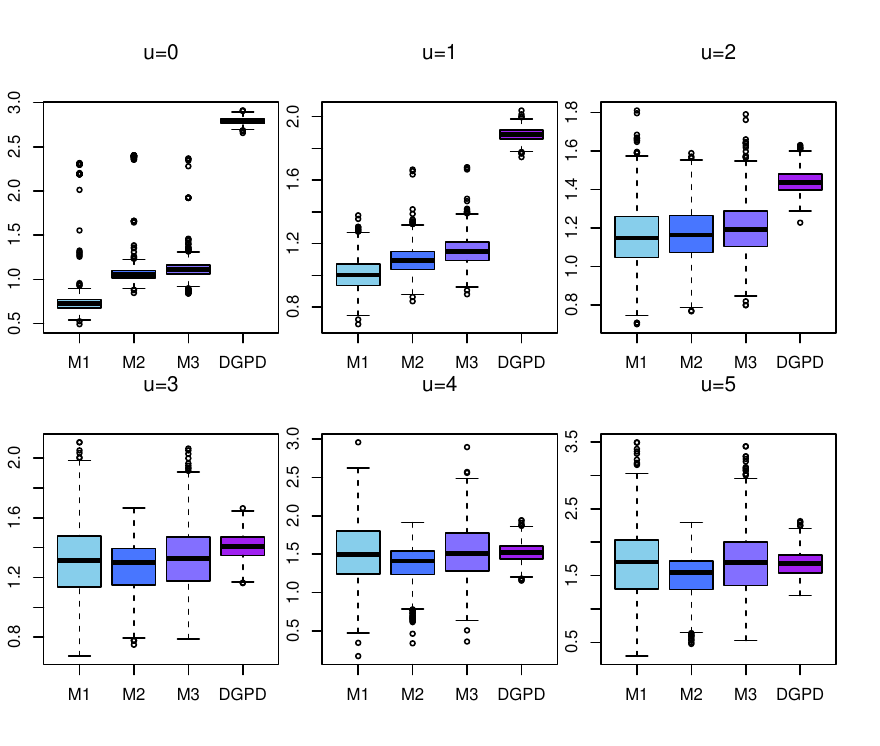}}}\\
\subfloat[$\xi$]{%
\resizebox*{6cm}{!}{\includegraphics{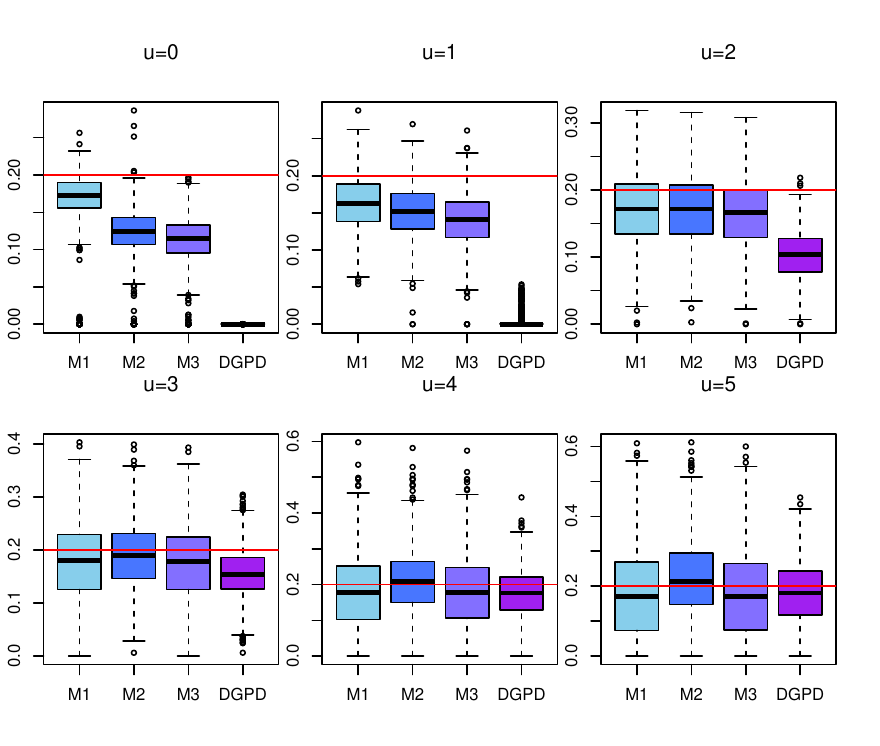}}}
\caption{Boxplots of estimated parameters of DEGPD models $M_1$, $M_2$ and $M_3$ at different threshold values $u=0, 1, 2, 3, 4, 5$. The data are generated from discrete Generalized extreme value distribution by
using the following parameter setting $\mu= 2, \beta = 1$ and $\xi = 1/20$. Boxplots are based on $n=2000$ sample points with $10^3$ independent replicates, and horizontal red lines in panel (c) represent true values of the tail index.}
	\label{fig:GEVsimtail}
\end{figure}

\subsection{Tail stability with low threshold}
To observe the right tail behavior of $M_1$, $M_2$, and $M_3$ at a low threshold, we simulate 1000 heavy-tailed samples of size $n=2000$ from discrete generalized extreme value distribution by using the following parameter setting $\mu=2$, $\beta=1$ and $\xi=1/20$. Here, the $\mu=2$ is chosen to generate only non-negative integer values in the samples. 

\begin{figure}
\centering
\subfloat[]{%
\resizebox*{14cm}{!}{\includegraphics{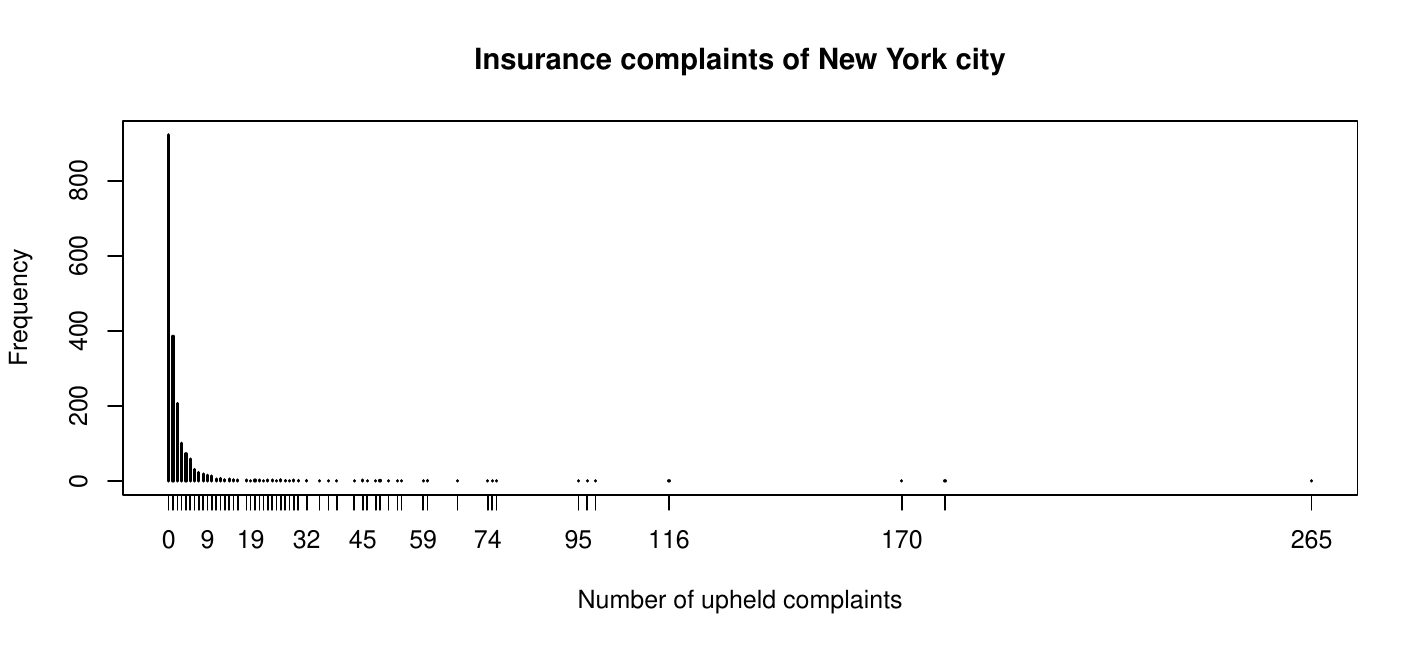}}}\hspace{5pt}\\
\subfloat[]{%
\resizebox*{7cm}{!}{\includegraphics{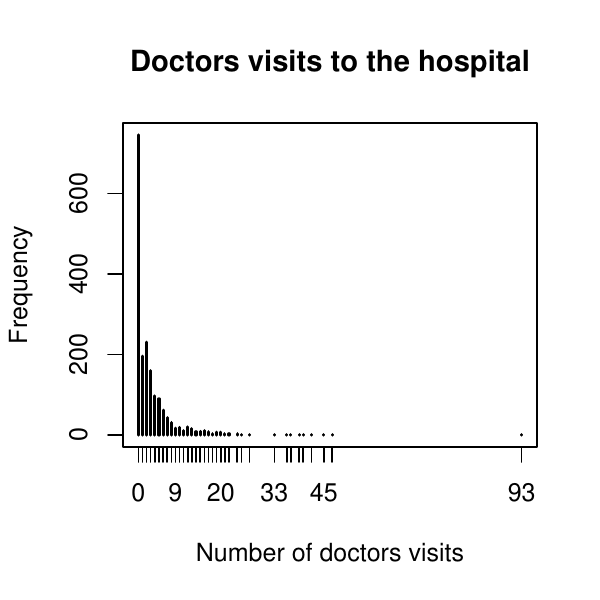}}}
\subfloat[]{%
\resizebox*{7cm}{!}{\includegraphics{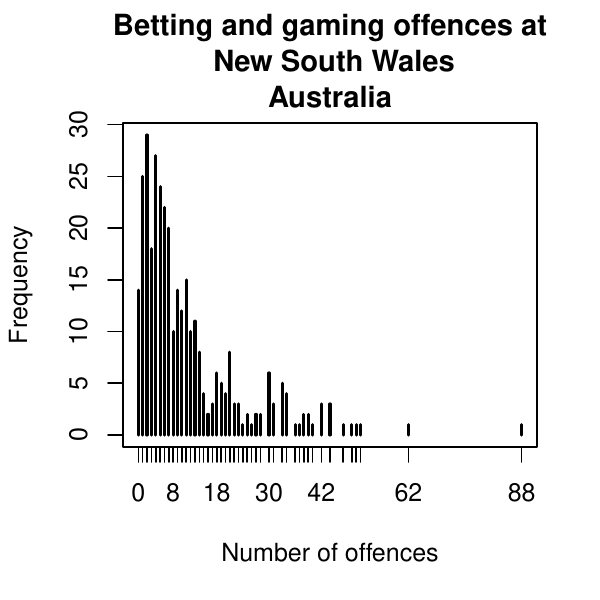}}}
  \caption{Frequency plot of the real data used for this study: (Top) number of upheld complaints of Automobile Insurance Company Complaints of New York City, (bottom left) number of doctor visits, and (bottom right) number of betting and gaming offenses in New Wales, Australia.}
    \label{fig:datahist}
\end{figure}

    
    

In Figure \ref{fig:GEVsimtail}, the parameter estimates for the DGPD and models ($M_1$, $M_2$, and $M_3$) are depicted against discrete thresholds $u=1, 2, 3, 4, 5$. Notably, the tail estimate exhibits a slight underestimation at lower thresholds but converges towards the true value  $\xi$  with increasing thresholds. The $M_1$, $M_2$, and $M_3$ models demonstrate a faster convergence of the tail estimate to the true value. In addition, a threshold of 
$u=5$ emerges as reasonable for modeling discrete exceedances using the DGPD, as the tail index estimates consistently approximate the true value of 
$\xi=1/20$. The $M_1$, $M_2$, and $M_3$ models confirm this since they all simplify the DGPD model above this threshold $u=5$. As a result, $\kappa$ tend to 1 for $M_1$, $M_3$ and approaches to 0 for $M_2$.

While all $M_1$, $M_2$, and $M_3$ models  effectively 
 reduce the bias in the tail index estimation compared to the DGPD model, particularly at lower thresholds; the $M_2$ model has a smaller bias, as shown in Figure \ref{fig:GEVsimtail}. In addition, the proposed DEGPD models are capable of describing the tail when the mode of the discrete exceedances is positive, i.e., when the threshold is below the mode of the distribution. In that case, however, the tail index is slightly underestimated.

\section{Real data applications} \label{realap}

This section shows how the proposed DEGPD and ZIDEPD models are useful in distinct real scenarios. Firstly, when the objective is to model the data with a moderate number of zeros and a heavy-tailed distribution, the DEGPD model emerges as a superior choice compared to existing models given in the motivating example. Secondly, in situations where the focus is on capturing excessive zeros and heavy-tailed behavior, the ZIDEGPD model stands out as the preferred option. Lastly, the flexibility of the DEGPD model is highlighted in cases involving low thresholds, proving advantageous when interpretative thresholds are not sufficiently high for the DGDP or when threshold selection poses a challenge. This versatility makes the DEGPD model particularly valuable in scenarios where threshold selection could be more complex. We consider three real data sets to test the proposed models in different scenarios. The first data set is about the Automobile Insurance Company Complaint Rankings (AICCR) of New York city \footnote{\href{ https://www.ny.gov/programs/open-ny.}{ https://www.ny.gov/programs/open-ny.}}.
 The second data set is related to the doctors visits (DV) in the hospital, which is taken from the \texttt{zic} R package. This data contains an excessive number of zeros and has a heavier tail, which is more appropriate for testing ZIDEGPD.
The third data set is related to the betting and gaming offenses in New Wales, Australia\footnote{\href{ https://www.bocsar.nsw.gov.au/}{ https://www.bocsar.nsw.gov.au/}} to test the flexibility of DEGPD at low thresholds. Figure \ref{fig:datahist} shows the frequency plots of all three data sets. Models fitted to the considered data sets are discussed in the following subsections.

 \subsection{DEGPD models}
\begin{figure}
\centering
\subfloat[DEGPD ($M_1$)]{%
\resizebox*{4.6cm}{!}{\includegraphics{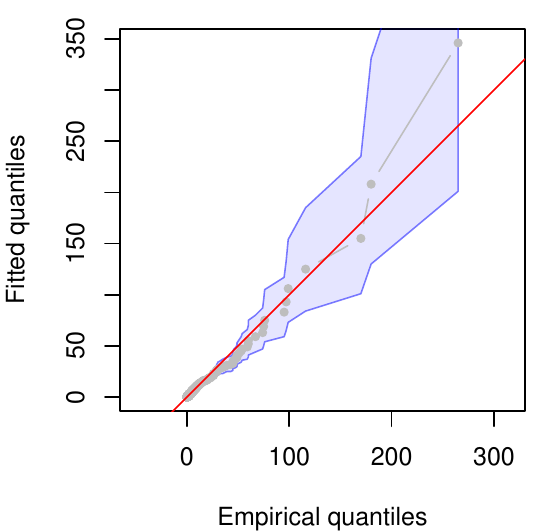}}}\hspace{5pt}
\subfloat[DEGPD ($M_2$)]{%
\resizebox*{4.6cm}{!}{\includegraphics{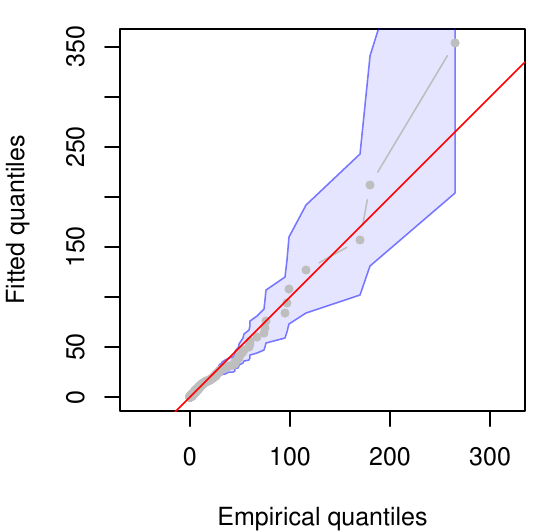}}}
\subfloat[DEGPD ($M_3$)]{%
\resizebox*{4.6cm}{!}{\includegraphics{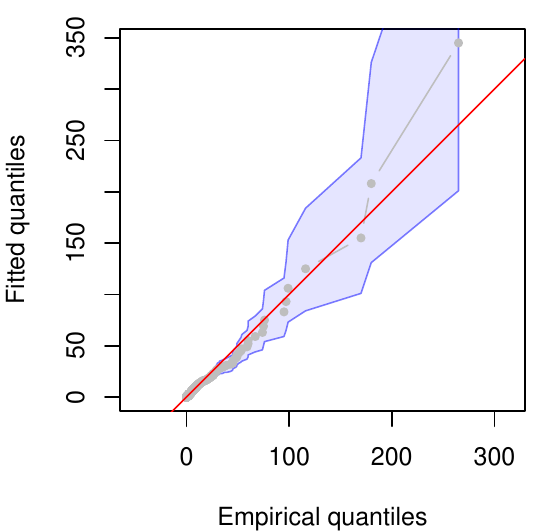}}}\\
\subfloat[DEGPD ($M_1$)]{%
\resizebox*{4.6cm}{!}{\includegraphics{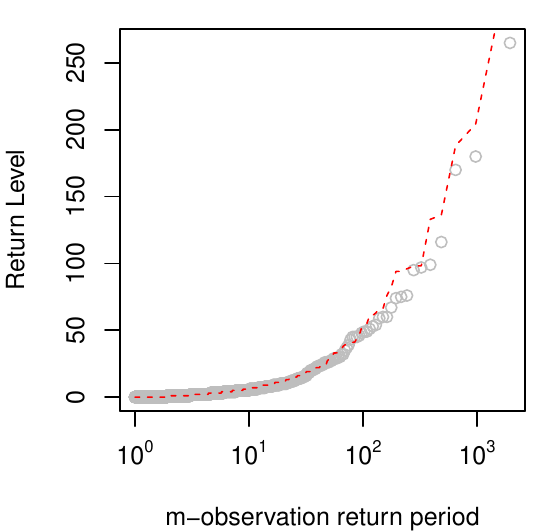}}}\hspace{5pt}
\subfloat[DEGPD ($M_2$)]{%
\resizebox*{4.6cm}{!}{\includegraphics{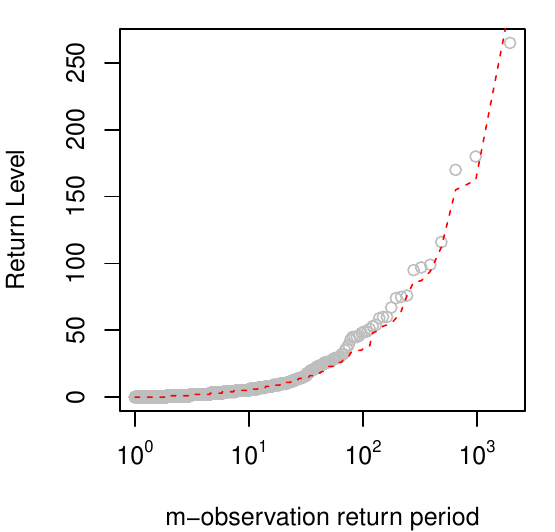}}}
\subfloat[DEGPD ($M_3$)]{%
\resizebox*{4.6cm}{!}{\includegraphics{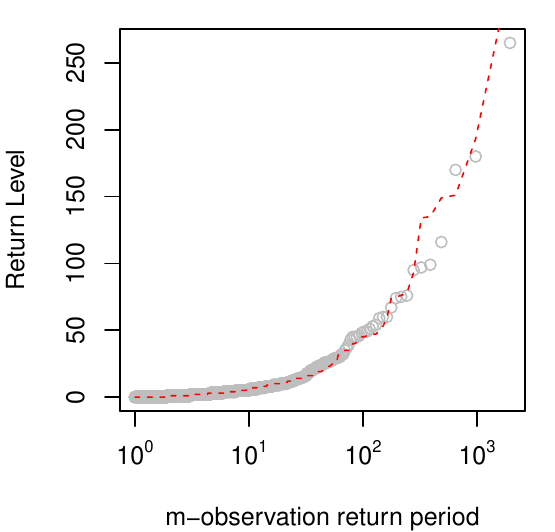}}}
  \caption{The Q-Q plots for a DEGPD $M_1$ $M_2$ and $M_3$ fitted to AICCR data (top). The m-observation return level plot of the fitted DEGPD $M_1$ $M_2$ and $M_3$ to
AICCR data (bottom). The shaded blue area in the top panel denotes the 95\%  bootstrap confidence intervals.  }
	\label{fig:fiting}
\end{figure}

The DEGPD models (represented by $M_1$, $M_2$, and $M_3$) were fitted to the AICCR dataset, and the results are summarized in Table \ref{tab:degpd_result}. All three models are evaluated using the Bayesian Information Criterion (BIC)
\[
\text{BIC} = -2 \cdot \ln(L) + k \cdot \ln(n),
\]
where $L$ is the maximum likelihood, $k$ represents the number of parameters in the model, and $n$ is the total number of observations in the observed set. Based on the BIC given in Table~\ref{tab:degpd_result}, model $M_2$ shows superior performance for the AICCR dataset across all models, \answerRrR{but higher variability is observed in parameter $\kappa$ when compared $M_1$ and $M_3$.} Despite variations in model performance, all DEGPD models exhibited satisfactory fits. Notably, DEGPD outperformed for datasets with a moderate number of zeros. In addition, the tail index estimated by all three models is similar (i.e., 0.73), while the bulk parameter $\kappa$ varies slightly. This clearly indicates that the proposed models show flexibility in the bulk part of distribution without compromising tail behavior. In contrast, existing models (see motivating examples) fail to capture extreme tail behavior when focused on the bulk and extreme tail of the distribution in the same operation. Additionally, Figure \ref{fig:fiting} (top) illustrated the Q-Q plots of the DEGPD fitted models, confirming the excellent fit of the models. A discrete Kolmogorov–Smirnov goodness-of-fit test from package \texttt{dgof} applied to the AICCR data set for all three models confirms an appropriate fit by its p-values \answerRR{(computed by Monte Carlo simulation)} given in the last column of Table~\ref{tab:degpd_result} \citep{arnold2011nonparametric}.
Further, m-observation return level plots in Figure~\ref{fig:fiting} (bottom) demonstrate a good fit regardless of the return period.
\begin{table}[t]
    \centering
\caption{Estimated parameters of DEGPD for $M_1, M_2$ and $M_3$ fitted to insurance claims data of New York City by maximum likelihood method. The 95\% bootstrap confidence bands are provided in square brackets, and standard errors are given in parentheses.
}
\begin{tabular}{ p{1cm} p{2.5cm} p{2.5cm} p{2cm} p{2cm} p{2cm}  }
\toprule
 \multicolumn{6}{c}{\textbf{DEGPD}} \\
 \hline
 &$\kappa$ &$\beta$ &$\xi$ &BIC& KS-test\\
 \hline
$M_1$&1.41 (0.36) & 0.8 (0.20) & 0.73 (0.05) &   7307.65 & 0.91\\
&$[ 1.01, 2.50]$&[0.40, 1.23]&[0.62, 0.84]\\
$M_2$&1.93 (1.12) & 0.82 (0.16) & 0.73 (0.05)&  $\textbf{7307.33}$ &  0.92\\
&$[0.08, 4.38]$&[0.55, 1.19]&[0.62, 0.84]\\
$M_3$&1.57 (0.41) & 0.88 (0.16) & 0.73 (0.06) &  7307.6 & 0.89\\
&$[1.01, 2.55]$&[0.59, 1.23]&[0.62, 0.83]\\
\bottomrule
\end{tabular}
\label{tab:degpd_result}
\end{table}

 \subsection{ZIEGPD}
\begin{figure}
\centering
\subfloat[ZIDEGPD ($M_1$)]{%
\resizebox*{4cm}{!}{\includegraphics{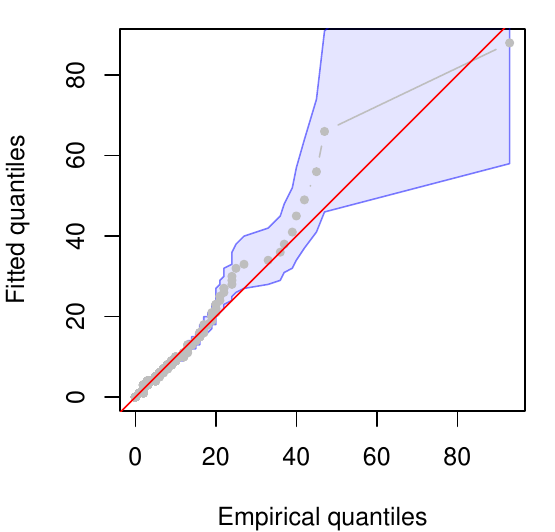}}}\hspace{5pt}
\subfloat[ZIDEGPD ($M_2$)]{%
\resizebox*{4cm}{!}{\includegraphics{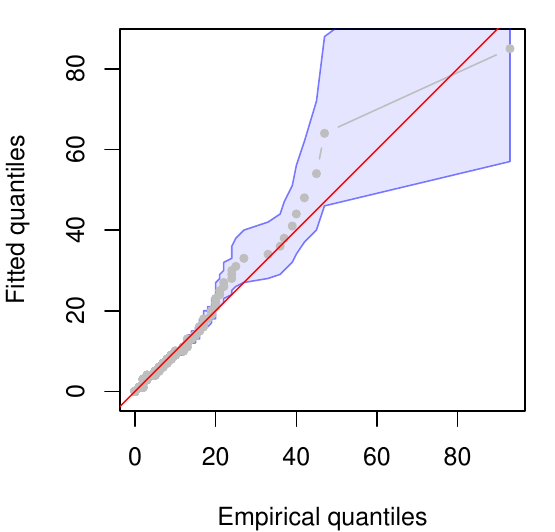}}}
\subfloat[ZIDEGPD ($M_3$)]{%
\resizebox*{4cm}{!}{\includegraphics{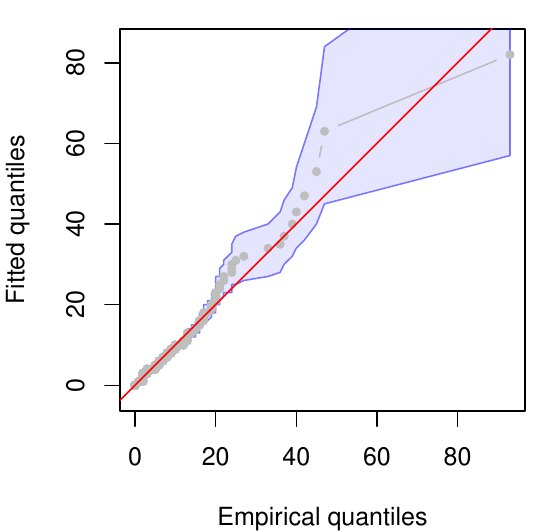}}}\\
\subfloat[ZIDEGPD ($M_1$)]{%
\resizebox*{4cm}{!}{\includegraphics{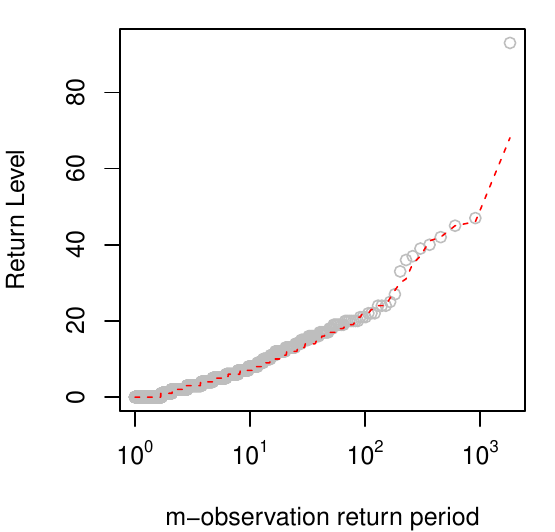}}}\hspace{5pt}
\subfloat[ZIDEGPD ($M_2$)]{%
\resizebox*{4cm}{!}{\includegraphics{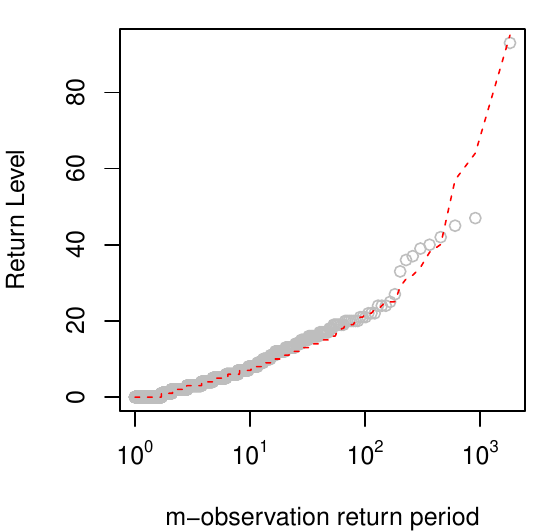}}}
\subfloat[ZIDEGPD ($M_3$)]{%
\resizebox*{4cm}{!}{\includegraphics{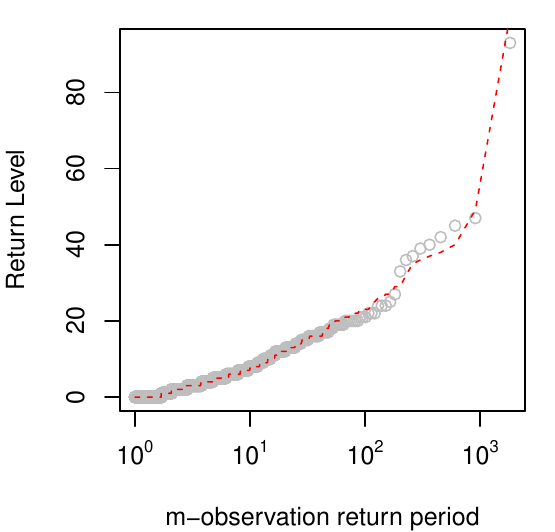}}}
 \caption{The Q-Q plots for a ZIDEGPD $M_1$ $M_2$ and $M_3$ fitted to DV data (top). The m-observation return level plot of the fitted ZIDEGPD $M_1$ $M_2$ and $M_3$ to
DV data (bottom). The shaded blue area in the top panel denotes the 95\%  bootstrap confidence intervals.}
	\label{fig:ZIGEPD-real}
\end{figure}

We assess the performance of the proposed ZIDEGPD models ($M_1$, $M_2$, and $M_3$) fitted to the DV dataset. The results of these fitted models are presented in Table~\ref{tab:zidegpd}. One notable observation is that the proportion of zeros, represented by the parameter $\pi$, is accurately estimated across all models. The estimated values of $\pi$ are very close to the true proportion of zeros in the dataset, demonstrating the ZIDEGPD model's ability to correctly capture the zero inflation in the data. There is variation across the models regarding the bulk part parameter $\kappa$. This variability arises because each model separately estimates the effect of zero values through the parameter $\pi$, which somehow influences the distribution of the bulk part of the data. 
The tail index parameter, $\xi$, remains relatively consistent across all models. This suggests that despite differences in bulk part estimation, the tail behavior of the data, captured by $\xi$, is similarly modeled in all cases. \answerRRR{Furthermore, the BIC value of $M_3$ indicates it performs slightly better than $M_1$ and $M_2$.}
Figure~\ref{fig:ZIGEPD-real} (top) presents the Q-Q plots for the ZIDEGPD fitted models, illustrating the best fit of models ($M_1$, $M_2$, and $M_3$) to the zero-inflated heavy tail data. 
\begin{table}[t]
    \centering
\caption{Estimated parameters of ZIDEGPD for $M_1, M_2$ and $M_3$ when applied to doctor visits data by the Maximum likelihood method. The 95\% bootstrap confidence bands are provided in square brackets,  \answerRrR{and standard errors are given in parentheses.}
}
\begin{tabular}{ p{0.7cm} p{2cm} p{2cm} p{2cm} p{2cm} p{1.5cm}  p{1.5cm}}
\toprule
 \multicolumn{7}{c}{\textbf{ZIDEGPD}} \\
\hline
 &$\pi$ &$\kappa$ &$\beta$ &$\xi$  &BIC & KS-test \\
\hline
$\text{M}_1$&0.38 (0.02) & 4.35 (2.93) & 1.22 (0.45)& 0.40 (0.06) &   3865.04 &0.70 \\
&$[ 0.32, 0.41]$&[1.98, 11.90]&[0.55, 2.25]&[0.27, 0.50]\\
$\text{M}_2$&0.37 (0.03)&9.71 (4.22) &1.66 (0.34)& 0.38 (0.05) & 3864.80 &0.70\\
&$[0.30, 0.41]$&[3.60, 19.68]&[1.16, 2.45]&[0.26, 0.47]\\
$\text{M}_3$&0.38 (0.02)& 5.29 (2.38)& 1.76 (0.35)& 0.37 (0.05)&  $\textbf{3864.44}$ &0.71\\
&$[0.33, 0.41]$&[2.63, 12.35]&[1.16, 2.57]&[0.25, 0.46]\\
\bottomrule
\end{tabular}
\label{tab:zidegpd}
\end{table}

These plots visually confirm that the observed quantiles closely match the expected quantiles under the fitted distributions, indicating a robust performance of the models. At the same time, this is not seen for existing models (see Figure~\ref{fig:existmodel}). A discrete Kolmogorov–Smirnov (K-S) goodness-of-fit test was conducted on the DV dataset for all three models to substantiate the adequacy of the model's further fits. 
The results of this test in terms of $p$-values \answerRR{(computed by Monte Carlo simulation)} are reflected in the last column of Table~\ref{tab:zidegpd} to support the appropriateness of the fitted models. High $p$-values suggest that there is no significant difference between the empirical distribution of the data and the theoretical distribution specified by the ZIDEGPD models, thereby reinforcing the validity of our proposed models. Additionally, the m-observation return level plots depicted in Figure~\ref{fig:ZIGEPD-real} (bottom) reveal a strong fit for the ZIDEGPD models across various return periods. These plots illustrate the expected maximum values over different time intervals, indicating that the models reliably capture the extreme value behavior of the data regardless of the return period considered. Thus, the ZIDEGPD models demonstrate strong performance in estimating the proportion of zeros and provide consistent tail index estimates.

\begin{table}[t]
    \centering
\caption{Estimated parameters and 95\% bootstrap confidence intervals of DEGPD ($M_3$) and DGPD when fitted to the exceedances of betting and gaming offenses
above the fixed low threshold values 1 and 3. \answerRrR{Bootstrap confidence
bands are provided in square brackets, and standard errors are given in parentheses.}  }
\begin{tabular}{ p{1.5cm} p{3.5cm} p{2cm} p{3.5cm} p{2cm} }
\toprule
 \multicolumn{5}{c}{${u=1}$} \\
 \hline
 \textbf{Model}&$\kappa$ &$\beta$ &$\xi$ &BIC \\
 \hline
 $\text{M}_1$&1.09 (0.14) & 9.61 (1.58) & 0.11 (0.07) &   \textbf{2265.24}  \\
&$[ 0.89, 1.40]$&[5.43, 12.67]&[1.60$\times 10^{-6}$, 0.26]\\
$\text{M}_2$&0.01 (0.75) & 10.6 (1.51) & 0.07 (0.08) &   2265.75  \\
&$[2.89\times10^{-6}, 2.83]$&[5.86, 11.98]&[1.36$\times 10^{-6}$, 0.29]\\
$\text{M}_3$&1.15 (0.35) & 9.66 (2.69) & 0.11 (0.08)   & 2265.34  \\
&$[0.13, 1.73]$&[6.62, 18.78]&[1.05$\times 10^{-6}$, 0.28]\\
$\text{DGPD}$&- & 10.62 (1.53)&0.06 (0.08)   &   2267.96 \\
&-&[8.88, 12.35]&[1.05$\times 10^{-2}$, 0.18]\\
 \hline
 \multicolumn{5}{c}{${u=3}$} \\
 \hline
 &$\kappa$ &$\beta$ &$\xi$ &BIC \\
 \hline
$\text{M}_1$&1.10 (0.17)& 9.27 (2.00) &0.13 (0.10)   &  \textbf{1895.74}  \\
&$[ 0.88, 1.52]$&[5.75, 13.02]&[1.32$\times 10^{-6}$, 0.35]\\
$\text{M}_2$&0.01 (0.73) & 10.48 (1.56) & 0.08 (0.09)&   1896.25  \\
&$[8.19\times10^{-7}, 2.89]$&[5.35, 12.13]&[1.58$\times 10^{-6}$, 0.35]\\
$\text{M}_3$&1.10 (0.40)& 9.80 (3.10) & 0.11 (0.10)&  1896.12 \\
&$[0.12, 1.87]$&[5.69, 18.92]&[1.07$\times 10^{-6}$, 0.35]\\
$\text{DGPD}$&- & 10.49 (1.57) &0.08 (0.10)&  1897.63  \\
&-&[8.55, 12.43]&[1.06$\times 10^{-2}$, 0.22]\\
\bottomrule
\end{tabular}
\label{tab:betting}
\end{table}

\subsection{DEGPD at low threshold}
\begin{figure}
\centering
\subfloat[DEGPD ($M_1$) when $u=1$]{%
\resizebox*{7cm}{!}{\includegraphics{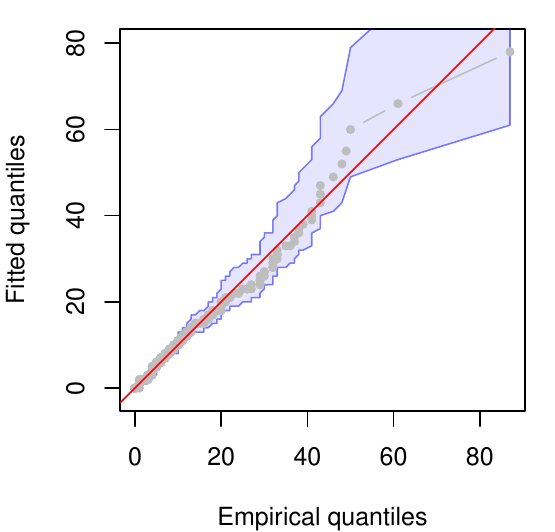}}}\hspace{5pt}
\subfloat[DEGPD ($M_1$) when $u=3$]{%
\resizebox*{7cm}{!}{\includegraphics{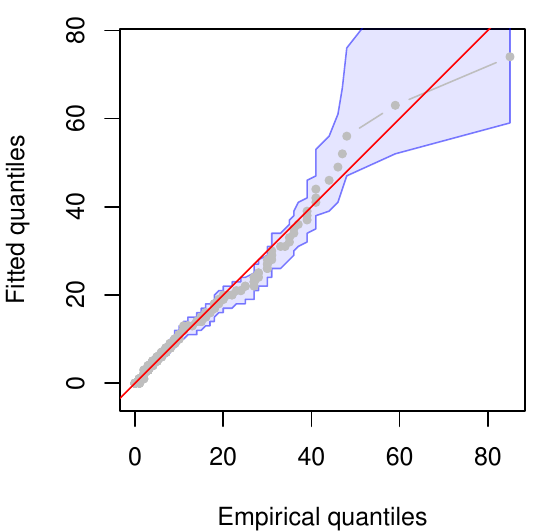}}}\\
\subfloat[DEGPD ($M_1$) when $u=1$]{%
\resizebox*{7cm}{!}{\includegraphics{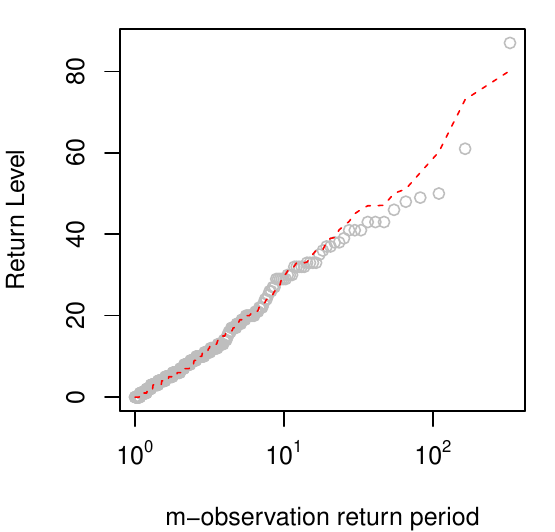}}}
\subfloat[DEGPD ($M_1$) when $u=3$]{%
\resizebox*{7cm}{!}{\includegraphics{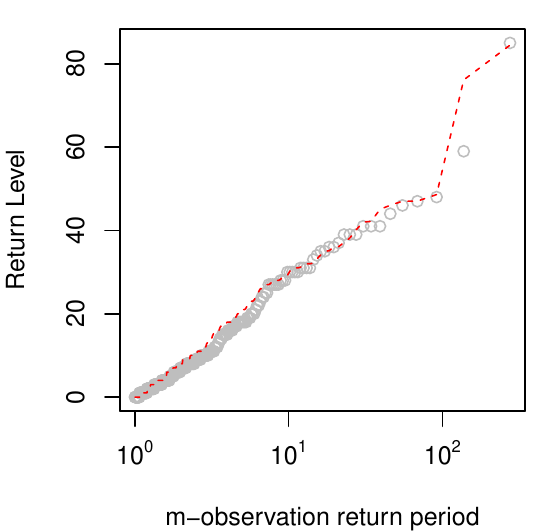}}}\hspace{5pt}
 \caption{(Left) Quantile-quantile plot and, (right) m-observation return level plot of the fitted \answerR{DEGPD ($M_1$)} model with low threshold values $u=1$ and 3.}
	\label{fig:Den_RL}
\end{figure}

To evaluate the performance of the DEGPD models at low threshold exceedances, Table~\ref{tab:betting} summarizes the parameter estimates and the BIC for the DEGPD models ($M_1$, $M_2$, and $M_3$) and DGPD. These models are specifically fitted to the exceedances of betting and gaming offenses above the fixed low threshold values 1 and 3. These threshold values are selected corresponding to the 10\% and 20\% quantiles of the data. According to the BIC, all DEGPD models outperform the DGPD, which is expected due to the threshold being too low for the DGPD model to be appropriate. With the DGPD model, the tail index estimate is underestimated at a lower level $\hat{\xi}=0.06 (1.05 \times 10^{-2}, 0.18)$ when the chosen threshold is very low, i.e., 1. It is consistent with the simulations (see Figure \ref{fig:GEVsimtail}), indicating that the tail index of DGPD is underestimated when the threshold is too low. For DEGPD ($M_1$) and DEGPD ($M_3$), the tail indexes are less underestimated, but for DEGPD ($M_2$), the tail index estimation is almost the same as DGPD when the chosen threshold is 3, a bit higher than low. \answerR{The BIC in Table \ref{tab:betting} favors the DEGPD($M_1$) model}. To observe what happens with high threshold selection, we fitted all models reported in Table \ref{tab:betting} with a threshold equal to 7. The results are not reported here. From non-reported results,  we observe that the tail indexes for DEGPD($M_1$) and DEGPD($M_3$) were underestimated, DGPD was performing better for exceedances over a high threshold, and the tail parameter of DEGPD($M_2$) stayed very near to DGPD tail index. Figure \ref{fig:Den_RL} shows the Q-Q and m-observation return level
plots for \answerR{DEGPD($M_1$)} when fitted model at low threshold values 1 and 3. The plots also confirm that the model is a good fit.

\section{Discussion and conclusion}\label{conl}
This paper proposes flexible versions of DEGPD, especially suitable for dealing with the entire range of count data, a large number of zeros, and conditional exceedances over a low threshold. We presented two new families of distributions that show flexibility in the above-discussed cases: one based on the truncated Normal distribution, another based on truncated \answer{Beta} distribution, and an existing one based on power law that has already been discussed in \citet{ahmad2024extended}. Consider distributions extending the DEGPD and ZIDEGPD models introduced in \citet{ahmad2024extended} by adding new extended models for which the distributions are flexible with nonnegative integers and zero excessiveness. 

Concerning the inference, the MLE approach performs adequately for both DEGPD and ZIDEGPD when tried for different combinations of parameters and sample sizes. In fact, DEGPD are considered to be more adequate for estimating the tail index when using a low threshold, as reported in Figure \ref{fig:GEVsimtail}. Using a low threshold along with DEGPD can improve the estimation of high quantiles, regardless of whether one is interested in extreme values. 

In application to AICCR, we found that the DEGPD fit adequately. The ZIDEPD also shows a good fit and accurately models the proportion of zero and tail behavior in DV data. It is also flexible when the data contains too many zeros and the non-zero integers have a Gamma-type distribution with a longer tail. On the other hand, when applying DEGPD to betting and gaming offenses in New South Wales, we found that the model fitted adequately when the low threshold was chosen to be less than 3.
In this manner, both the bulk part and the right tail were effectively modeled. As we have seen, the proposed models fit satisfactorily in all scenarios even when the very low threshold is chosen for both the simulation study and real data. The approach introduced in this article is simpler and does not require fixing a high enough threshold to achieve a better overall fit. Future research could explore the theoretical implications of the proposed and existing threshold approaches to determine which is more advantageous in different scenarios.

The proposed models offer a way to assess the suitability of a threshold to apply the DGPD to model discrete exceedances above that threshold. Specifically, if the shape parameter in $M_2$ is near 0, it suggests that the asymptotic approximation is valid and that the DGPD is also appropriate for the chosen threshold. Similarly, for the $M_1$ model, a shape parameter of 1 indicates that the DGPD is again suitable.

Furthermore, the DEGPD models could be used to develop methods for threshold selection in integer-valued data. In contrast to the traditional threshold-relying model discussed in \citet{hitz_davis_samorodnitsky_2024}, where selecting a sufficiently high threshold is crucial for the asymptotic approximation to be valid, extended models offer greater flexibility near the threshold. Therefore, threshold selection is less critical in extended models, as their probability mass functions can adapt more readily to variations near the threshold.

To conclude, DEGPD models are expected to be particularly useful for analyzing real data involving non-negative discrete random variables in three specific \answerR{contexts.} First, DEGPD models can be effectively applied to datasets consisting of non-negative integer values throughout their entire range. This includes situations where the data naturally takes on integer values, such as counts of events or occurrences. The DEGPD models are capable of capturing the distribution characteristics of these integer-valued data, including their variability and distribution shape. Second, for datasets where non-negative integer values are prevalent but also have a significant number of zero observations (often referred to as "excessive zeros"), DEGPD models with their ZI version can handle this data effectively. This is particularly important in fields like environmental data or medical statistics, where zeros might occur more frequently than other values. DEGPD models are designed to account for such zero inflation while still modeling the non-zero counts accurately. Third, DEGPD models are also suitable for analyzing discrete data where interest lies in exceedances above a specified threshold. This scenario is common in extreme value analysis, such as modeling rare and extreme events (e.g., high precipitation levels or extreme temperature spike counts). DEGPD models provide a framework for modeling these exceedances while accounting for the discrete nature of the data.



\section*{Disclosure statement}

The authors declare no conflict of interest.

\section*{Funding}

This study is not receiving any specific funding.

\section*{Notes on contributor(s)}

Both authors equally contributed to this research.

\section*{Data and code availability}
The data and codes prepared for these models are available online in the GitHub repository.  \href{ https://github.com/touqeerahmadunipd/New_flexible_DEGPD}{ https://github.com/touqeerahmadunipd/New\_flexible\_DEGPD}.

\bibliographystyle{plainnat}
\bibliography{bibliography}

\begin{thebibliography}{23}
\providecommand{\natexlab}[1]{#1}
\providecommand{\url}[1]{\texttt{#1}}
\expandafter\ifx\csname urlstyle\endcsname\relax
  \providecommand{\doi}[1]{doi: #1}\else
  \providecommand{\doi}{doi: \begingroup \urlstyle{rm}\Url}\fi

\bibitem[Ahmad et~al.(2024)Ahmad, Gaetan, and Naveau]{ahmad2024extended}
Touqeer Ahmad, Carlo Gaetan, and Philippe Naveau.
\newblock An extended generalized pareto regression model for count data.
\newblock \emph{Statistical Modelling}, page 1471082X241266729, 2024.

\bibitem[Arnold and Emerson(2011)]{arnold2011nonparametric}
Taylor~B Arnold and John~W Emerson.
\newblock Nonparametric goodness-of-fit tests for discrete null distributions.
\newblock \emph{R Journal}, 3\penalty0 (2), 2011.

\bibitem[Balkema and De~Haan(1974)]{balkema:de_haan:1974}
August~A Balkema and Laurens De~Haan.
\newblock Residual life time at great age.
\newblock \emph{The Annals of Probability}, 2:\penalty0 792--804, 1974.

\bibitem[Buddana and Kozubowski(2014)]{buddana2014discrete}
Amrutha Buddana and Tomasz~J Kozubowski.
\newblock Discrete pareto distributions.
\newblock \emph{Economic Quality Control}, 29\penalty0 (2):\penalty0 143--156, 2014.

\bibitem[Carreau and Bengio(2009)]{carreau2009hybrid}
Julie Carreau and Yoshua Bengio.
\newblock A hybrid pareto model for asymmetric fat-tailed data: the univariate case.
\newblock \emph{Extremes}, 12\penalty0 (1):\penalty0 53--76, 2009.

\bibitem[Carrer and Gaetan(2022)]{carrer2022distributional}
No{\'e}mie~Le Carrer and Carlo Gaetan.
\newblock Distributional regression models for {E}xtended {G}eneralized {P}areto distributions.
\newblock \emph{arXiv preprint arXiv:2209.04660}, 2022.

\bibitem[Coles(2001)]{coles2001introduction}
Stuart Coles.
\newblock \emph{An Introduction to Statistical Modeling of Extreme Values}.
\newblock Springer, New York, 2001.

\bibitem[Daouia et~al.(2023)Daouia, Stupfler, and Usseglio-Carleve]{daouia:2023}
Abdelaati Daouia, Gilles Stupfler, and Antoine Usseglio-Carleve.
\newblock Extreme value modelling of {SARS}-{CoV}-2 community transmission using discrete generalized pareto distributions.
\newblock \emph{Royal Society Open Science}, 10, March 2023.
\newblock \doi{10.1098/rsos.220977}.
\newblock URL \url{https://doi.org/10.1098/rsos.220977}.

\bibitem[Davison and Smith(1990)]{davison1990models}
Anthony~C Davison and Richard~L Smith.
\newblock Models for exceedances over high thresholds.
\newblock \emph{Journal of the Royal Statistical Society Series B: Statistical Methodology}, 52\penalty0 (3):\penalty0 393--425, 1990.

\bibitem[Frigessi et~al.(2002)Frigessi, Haug, and Rue]{frigessi2002dynamic}
Arnoldo Frigessi, Ola Haug, and H{\aa}vard Rue.
\newblock A dynamic mixture model for unsupervised tail estimation without threshold selection.
\newblock \emph{Extremes}, 5\penalty0 (3):\penalty0 219--235, 2002.

\bibitem[Gamet and Jalbert(2022)]{gamet2022}
Phil\'emon Gamet and Jonathan Jalbert.
\newblock A flexible extended generalized {P}areto distribution for tail estimation.
\newblock \emph{Environmetrics}, 33\penalty0 (6):\penalty0 e2744, 2022.

\bibitem[Hitz et~al.(2024)Hitz, Davis, and Samorodnitsky]{hitz_davis_samorodnitsky_2024}
Adrien~S. Hitz, Richard~A. Davis, and Gennady Samorodnitsky.
\newblock Discrete extremes.
\newblock \emph{Journal of Data Science}, 22\penalty0 (4):\penalty0 524--536, 2024.

\bibitem[Kozubowski et~al.(2015)Kozubowski, Panorska, and Forister]{kozubowski2015discrete}
Tomasz~J Kozubowski, Anna~K Panorska, and Matthew~L Forister.
\newblock A discrete truncated pareto distribution.
\newblock \emph{Statistical Methodology}, 26:\penalty0 135--150, 2015.

\bibitem[Krishna and Pundir(2009)]{krishna2009discrete}
Hare Krishna and Pramendra~Singh Pundir.
\newblock Discrete burr and discrete pareto distributions.
\newblock \emph{Statistical Methodology}, 6\penalty0 (2):\penalty0 177--188, 2009.

\bibitem[Lambert(1992)]{lambert1992zero}
Diane Lambert.
\newblock Zero-inflated poisson regression, with an application to defects in manufacturing.
\newblock \emph{Technometrics}, 34\penalty0 (1):\penalty0 1--14, 1992.

\bibitem[MacDonald et~al.(2011)MacDonald, Scarrott, Lee, Darlow, Reale, and Russell]{macdonald2011flexible}
A~MacDonald, Carl~John Scarrott, Dominic Lee, Brian Darlow, Marco Reale, and Glynn Russell.
\newblock A flexible extreme value mixture model.
\newblock \emph{Computational Statistics \& Data Analysis}, 55\penalty0 (6):\penalty0 2137--2157, 2011.

\bibitem[Naveau et~al.(2016)Naveau, Huser, Ribereau, and Hannart]{naveau2016modeling}
Philippe Naveau, Rapha{\"e}l Huser, Pierre Ribereau, and Alexis Hannart.
\newblock Modeling jointly low, moderate, and heavy rainfall intensities without a threshold selection.
\newblock \emph{Water Resources Research}, 52\penalty0 (4):\penalty0 2753--2769, 2016.

\bibitem[Papastathopoulos and Tawn(2013)]{papastathopoulos2013extended}
Ioannis Papastathopoulos and Jonathan~A Tawn.
\newblock Extended generalised pareto models for tail estimation.
\newblock \emph{Journal of Statistical Planning and Inference}, 143\penalty0 (1):\penalty0 131--143, 2013.

\bibitem[Pickands(1975)]{pickands1975statistical}
James Pickands.
\newblock Statistical inference using extreme order statistics.
\newblock \emph{the Annals of Statistics}, pages 119--131, 1975.

\bibitem[Ranjbar et~al.(2022)Ranjbar, Cantoni, Chavez-Demoulin, Marra, Radice, and Jaton]{ranjbar2022modelling}
Setareh Ranjbar, Eva Cantoni, Val{\'e}rie Chavez-Demoulin, Giampiero Marra, Rosalba Radice, and Katia Jaton.
\newblock Modelling the extremes of seasonal viruses and hospital congestion: The example of flu in a swiss hospital.
\newblock \emph{Journal of the Royal Statistical Society Series C: Applied Statistics}, 71\penalty0 (4):\penalty0 884--905, 2022.

\bibitem[Schwertman et~al.(2004)Schwertman, Owens, and Adnan]{schwertman2004simple}
Neil~C Schwertman, Margaret~Ann Owens, and Robiah Adnan.
\newblock A simple more general boxplot method for identifying outliers.
\newblock \emph{Computational statistics \& data analysis}, 47\penalty0 (1):\penalty0 165--174, 2004.

\bibitem[Stasinopoulos et~al.(2018)Stasinopoulos, Rigby, and Bastiani]{stasinopoulos2018gamlss}
Mikis~D Stasinopoulos, Robert~A Rigby, and Fernanda~De Bastiani.
\newblock Gamlss: A distributional regression approach.
\newblock \emph{Statistical Modelling}, 18\penalty0 (3-4):\penalty0 248--273, 2018.

\bibitem[Stein(2021)]{stein:2021}
Michael~L. Stein.
\newblock A parametric model for distributions with flexible behavior in both tails.
\newblock \emph{Environmetrics}, 32:\penalty0 e2658, 2021.

\end{thebibliography}

\newpage

\renewcommand{\thetable}{S.\arabic{table}}
\setcounter{table}{0}
\renewcommand{\thefigure}{S.\arabic{figure}}
\setcounter{figure}{0}

\renewcommand{\theequation}{S.\arabic{equation}}
\setcounter{equation}{0}
\renewcommand{\thesection}{S.\arabic{section}}
\setcounter{section}{0}
\setcounter{page}{1}

 \begin{center}
 	\section*{Supplementary material for \\ ``New flexible versions of extended generalized Pareto distribution for
count data''}
 \end{center}

\answerRrR{
\section{Theoretical result of EGPD and DEGPD Model 1}
Let $Z\geq 0$ be a continuous random variable. For $\mathcal{G}(\nu; \kappa)=\nu^{\kappa}$, $\kappa >0$, which is Model $M_1$ in the main document, the CDF of the EGPD model is written as
\begin{equation}\label{cdf-power}
H_Z(z) =
\begin{cases}
\left[ 1 - \left( 1 + \dfrac{\xi z}{\beta} \right)^{-\tfrac{1}{\xi}} \right]^{\kappa}, & \xi \neq 0, \\[1.2em]
\left[ 1 - \exp\!\left(-\dfrac{z}{\beta}\right) \right]^{\kappa}, & \xi = 0,
\end{cases}
\end{equation}
For $\xi>0$, it is easy to verify that $1-H_{Z}(z) \sim \kappa\left(\beta /\xi\right)^{1/\xi}  z^{-1/\xi}$, as $z\rightarrow \infty$.
Therefore, this three-parameter EGPD complies with EVT in
both tails, with $\kappa$ a shape parameter controlling the lower tail behavior, $\xi$ a shape parameter
controlling the upper tail behavior, and $\sigma$ an overall scale parameter affecting the whole
distribution from low to high quantiles.
}



\answerRrR{
\begin{theorem}[Regular variation of continuous EGPD]
Let $Z\sim \mathrm{EGPD}(\kappa,\beta,\xi)$ with $\kappa>0,\ \beta>0,\ \xi>0$ and have the CDF~\ref{cdf-power}. Then:
\begin{enumerate}
\item The survival function satisfies
\[
\overline H_Z(z)=1-(1-\omega(z))^{\kappa}\sim \kappa \omega(z), \qquad z\to\infty,
\]
where $\omega(z)=(1+\xi z/\beta)^{-1/\xi}$.
\item As $z\to\infty$,
\[
\omega(z)\sim \Big(\beta/{\xi}\Big)^{1/\xi} z^{-1/\xi}.
\]
Therefore
\[
\overline H_Z(z)\sim \kappa\Big({\beta}/{\xi}\Big)^{1/\xi} z^{-1/\xi},\qquad z\to\infty.
\]
\item Hence $\overline H_Z\in RV_{-1/\xi}$ and $H_Z$ belongs to the Fréchet maximum domain of attraction with index $\xi$, i.e.,
\(
H_Z\in \mathrm{MDA}_\xi.
\)
\end{enumerate}
\end{theorem}
\begin{proof}
(1) For small $\omega$, the generalized binomial expansion is $(1-\omega)^\kappa=1-\kappa \omega+o(\omega)$. As $z\to\infty$, $\omega(z)\to 0$, so
\[
\overline H_Z(z)=1-(1-\omega(z))^\kappa = 1 - (1 - \kappa \omega(z) + o(\omega(z))) = \kappa \omega(z)+o(\omega(z)).
\]
Thus $\overline H_Z(z)\sim \kappa \omega(z)$.
\\
(2) Rewrite $\omega(z)$ by factoring out the term $\xi z/\beta$:
\[
\omega(z)=\Big(1+\frac{\xi z}{\beta}\Big)^{-1/\xi}
=\Big(\frac{\xi z}{\beta}\Big)^{-1/\xi}\Big(1+\frac{\beta}{\xi z}\Big)^{-1/\xi}
=\Big(\frac{\beta}{\xi}\Big)^{1/\xi} z^{-1/\xi} \Big(1+\frac{\beta}{\xi z}\Big)^{-1/\xi}.
\]
Since $\big(1+\frac{\beta}{\xi z}\big)^{-1/\xi}\to1$ as $z\to\infty$, we obtain
$\omega(z)\sim \big(\frac{\beta}{\xi}\big)^{1/\xi}z^{-1/\xi}$.
Combining this with part (1) yields the asymptotic expansion for the survival function:
\[
\overline H_Z(z)\sim \kappa \omega(z) \sim \kappa\Big(\frac{\beta}{\xi}\Big)^{1/\xi} z^{-1/\xi}.
\]
(3) Combining (1) and (2) yields the asymptotic expansion.  
Thus $\overline H_Z(z) \sim c z^{-1/\xi}$ with $c = \kappa(\beta/\xi)^{1/\xi}$, so $\overline H_Z\in RV_{-1/\xi}$. By the Fisher--Tippett--Gnedenko theorem, $H_Z\in \mathrm{MDA}_\xi$ (Fréchet maximum domain of attraction) with extreme value index $\xi$, i.e., \(
H_Z\in \mathrm{MDA}_\xi.
\).
\end{proof}
}

\answerRrR{
\begin{theorem}[Discrete power law EGPD]
Let $Z\sim \mathrm{EGPD}(\kappa,\beta,\xi)$ with $\xi>0$ and define $Y=\lfloor Z\rfloor$. Then:
\begin{enumerate}
\item The discrete survival function for each integer $k \ge0$ is
\[
\overline F_Y(k):=\Pr(Y\ge k)=\Pr(Z\ge k)=\overline H_Z(k).
\]
\item As $k\to\infty$,
\[
\overline F_Y(k)\sim \kappa\Big(\tfrac{\beta}{\xi}\Big)^{1/\xi} k^{-1/\xi}.
\]
\item Therefore $\overline F_Y\in RV_{-1/\xi}$, i.e.\ the discrete EGPD preserves the same tail index as the continuous case, and $F_Y$ belongs to the discrete maximum domain of attraction with index $\xi$.
\end{enumerate}
\end{theorem}
}
\answerRrR{
\begin{proof}
For integer $k\ge0$, the floor is
\[
\{Y\ge k\}=\{\lfloor Z\rfloor\ge k\}=\{Z\ge k\},
\]
we have
\[
\overline F_Y(k)=\Pr(Y\ge k)=\Pr(\lfloor Z\rfloor\ge k)=\overline H_Z(k).
\]
From the continuous case, $\overline H_Z(z)\sim \kappa(\beta/\xi)^{1/\xi} z^{-1/\xi}$. Evaluating at $z=k$ gives
\[
\overline F_Y(k)\sim \kappa\Big(\tfrac{\beta}{\xi}\Big)^{1/\xi} k^{-1/\xi}.
\]
Thus $\overline F_Y(k)=k^{-1/\xi}L(k)$ with $L(k)\to \kappa(\beta/\xi)^{1/\xi}$ slowly varying, so $\overline F_Y\in RV_{-1/\xi}$. By discrete extreme value theory arguement given in~\citep{hitz_davis_samorodnitsky_2024}, this implies $F_Y\in \mathrm{MDA}_\xi$.
\end{proof}
}

\answerRrR{
\section{Theoretical results of EGPD and DEGPD Model 2}
Let $Z\geq 0$ be a continuous random variable. For
$\mathcal{G}(\nu; \kappa)=\frac{2}{2\Phi\left(\sqrt{\kappa}\right)-1} \Big[\Phi\left(\sqrt{\kappa}(\nu-1)\right)-\left\{1-\Phi(\sqrt{\kappa})\right\}\Big]$, the defined $\mathcal{G}(u)$ is the cdf of truncated normal distribution with precision $\kappa>0$ and $\Phi$ is the cdf of standard normal distribution. The CDF of the EGPD Model 2 is written as
\begin{equation}\label{cdf-tn}
   H_Z(z)=\frac{2}{2\Phi\left(\sqrt{\kappa}\right)-1} \Big[\Phi\left(\sqrt{\kappa}(G(z, \beta, \xi)-1)\right)-\left\{1-\Phi(\sqrt{\kappa})\right\}\Big],
\end{equation}
where $G(., \beta, \xi)$ is the CDF of GPD. The parameter $\kappa$ is a shape parameter controlling the lower tail behavior, $\xi$ is a shape parameter
controlling the upper tail behavior, and $\sigma$ an overall scale parameter affecting the whole
distribution from low to high quantiles.
}

\answerRrR{
\begin{theorem}[Regular variation of truncated normal mapped EGPD]
Let $Z\sim H_Z(\cdot;\kappa, \beta, \xi)$ with $\kappa>0,\ \beta>0,\ \xi>0$, where $H_Z$ is the truncated normal mapped EGPD given~\ref{cdf-tn}. Then:
\begin{enumerate}
\item The survival function satisfies
\[
\overline H_Z(z) \sim K\, \omega(z), \qquad z\to\infty,
\]
where $\omega(z)=(1+\xi z/\beta)^{-1/\xi}$ and 
$K = \dfrac{2\sqrt{\kappa}}{[2\Phi(\sqrt{\kappa})-1]\sqrt{2\pi}}$.
\item As $z\to\infty$,
\[
\omega(z)\sim \left(\frac{\beta}{\xi}\right)^{1/\xi} z^{-1/\xi}.
\]
Therefore
\[
\overline H_Z(z)\sim K\left(\frac{\beta}{\xi}\right)^{1/\xi} z^{-1/\xi},\qquad z\to\infty.
\]
\item Hence $\overline H_z\in RV_{-1/\xi}$ and $H_Z$ also belongs to the Fr\'echet maximum domain of attraction with index $\xi$, i.e.,
$
H_Z\in \mathrm{MDA}_\xi.
$
\end{enumerate}
\end{theorem}
}
\answerRrR{
\begin{proof}
Let us define:
\begin{itemize}
\item $A = 2\Phi(\sqrt{\kappa}) - 1$ (note $A > 0$ since $\kappa > 0$)
\item $B = 1 - \Phi(\sqrt{\kappa}) = \Phi(-\sqrt{\kappa})$
\item $\omega(z) = 1 - G(z) = \left(1 + \frac{\xi z}{\beta}\right)^{-1/\xi}$ (this is the GPD survival function)
\end{itemize}
Observe that $A = 1 - 2B$. The CDF becomes $
H_Z(z) = \frac{2}{A}\left[\Phi\big(\sqrt{\kappa}(G(z)-1)\big) - B\right].
$
\\
The survival function is expressed as
$
\overline H_Z(z) = 1 - H_Z(z) = 1 - \frac{2}{A}\Phi\big(\sqrt{\kappa}(G(z)-1)\big) + \frac{2B}{A}.
$
Note that $G(z) = 1 - \omega(z)$, so:
\[
\sqrt{\kappa}(G(z)-1) = -\sqrt{\kappa}\, \omega(z).
\]
Thus:
\[
\overline H_Z(z) = 1 + \frac{2B}{A} - \frac{2}{A}\Phi\big(-\sqrt{\kappa}\, \omega(z)\big).
\]
Since $A = 1 - 2B$, we have $1 + \frac{2B}{A} = \frac{A + 2B}{A} = \frac{1}{A}$. Therefore:
\begin{equation}\label{sfff}
\overline H_Z(z) = \frac{1}{A} - \frac{2}{A}\Phi\big(-\sqrt{\kappa}\, \omega(z)\big) = \frac{1}{A}\left[1 - 2\Phi\big(-\sqrt{\kappa}\, \omega(z)\big)\right]. 
\end{equation}
As $z \to \infty$, $\omega(z) \to 0$. Let $t = -\sqrt{\kappa}\, \omega(z) \to 0^{-}$.
\\
We use the Taylor expansion of $\Phi(t)$ around $t = 0$ as
$
\Phi(t) = \frac{1}{2} + \frac{t}{\sqrt{2\pi}} + O(t^3) \quad \text{as } t \to 0.
$
Substituting $t = -\sqrt{\kappa}\, \omega(z)$:
\[
\Phi\big(-\sqrt{\kappa}\, \omega(z)\big) = \frac{1}{2} - \frac{\sqrt{\kappa}\, \omega(z)}{\sqrt{2\pi}} + O(\omega(z)^3).
\]
Using equation~\eqref{sfff}
\[
\overline H_Z(z) = \frac{1}{A}\left[1 - 2\left(\frac{1}{2} - \frac{\sqrt{\kappa}\, \omega(z)}{\sqrt{2\pi}} + O(\omega(z)^3)\right)\right].
\]
After simplification, we get
\[
\overline H_Z(z) = \frac{1}{A} \cdot \frac{2\sqrt{\kappa}\, \omega(z)}{\sqrt{2\pi}} + O(\omega(z)^3) = \frac{2\sqrt{\kappa}}{A\sqrt{2\pi}}\, \omega(z) + O(\omega(z)^3).
\]
Therefore:
\[
\overline H_Z(z) \sim K\, \omega(z), \quad \text{where } K = \frac{2\sqrt{\kappa}}{A\sqrt{2\pi}} = \frac{2\sqrt{\kappa}}{[2\Phi(\sqrt{\kappa})-1]\sqrt{2\pi}}.
\]
This proves part (1).
\\
For the asymptotic of $\omega(z)$. We have
\[\omega(z)=\Big(1+\frac{\xi z}{\beta}\Big)^{-1/\xi}
=\Big(\frac{\xi z}{\beta}\Big)^{-1/\xi}\Big(1+\frac{\beta}{\xi z}\Big)^{-1/\xi}
=\Big(\frac{\beta}{\xi}\Big)^{1/\xi} z^{-1/\xi} \Big(1+\frac{\beta}{\xi z}\Big)^{-1/\xi}.\]
As $z \to \infty$, $\left(1 + \frac{\beta}{\xi z}\right)^{-1/\xi} \to 1$, so:
\[
\omega(z) \sim   \left(\frac{\beta}{\xi}\right)^{1/\xi} z^{-1/\xi}.
\]
This proves part (2).
\\
Combine proofs of (1) and (2) for the purpose of tail behavior
\[
\overline H_Z(z) \sim K\, \omega(z) \sim K\left(\frac{\beta}{\xi}\right)^{1/\xi} z^{-1/\xi}.
\]
Thus $\overline H_Z^{(2)}(z) = z^{-1/\xi} L(z)$ where $L(z) \to K(\beta/\xi)^{1/\xi}$ is a slowly varying function. Therefore $\overline H_Z \in RV_{-1/\xi}$.
\\
By the Fisher--Tippett--Gnedenko theorem, if $\overline H_Z \in RV_{-\frac{1}{\xi}}$ with $\xi > 0$, then $H_Z$ also belongs to the Fr\'echet maximum domain of attraction with extreme value index $\xi$. Indeed $H_Z \in \mathrm{MDA}_\xi$.
\\
This completes the proof of part (3).
\end{proof}
}

\answerRrR{
\begin{theorem}[Discrete truncated normal mapped EGPD]
Let $Z\sim H_Z(\cdot;\kappa, \beta, \xi)$ satisfy the previous theorem's assumptions (so $\overline H_Z(z)\sim K\,\omega(z)$ with $K>0$). Define $Y=\lfloor Z\rfloor$. Then:
\begin{enumerate}
\item The discrete survival function satisfies
\[
\overline F_Y(k)=\Pr(Y\ge k)\sim K\left(\frac{\beta}{\xi}\right)^{1/\xi} k^{-1/\xi},
\qquad k\to\infty.
\]
\item Hence $\overline F_Y\in RV_{-1/\xi}$ and $F_Y\in\mathrm{MDA}_\xi$.
\end{enumerate}
\end{theorem}
}
\answerRrR{
\begin{proof}
For integer $k\ge0$,
\[
\{Y\ge k\}=\{\lfloor Z\rfloor\ge k\}=\{Z\ge k\},
\]
so $\overline F_Y(k)=\Pr(Y\ge k)=\Pr(Z\ge k)=\overline H_Z(k)$. From the continuous result,
\[
\overline H_Z(z)\sim K\,\omega(z)=K\Big(1+\frac{\xi z}{\beta}\Big)^{-1/\xi},
\]
hence, evaluating at $z=k$,
\[
\overline F_Y(k)=\overline H_Z(k)\sim K\Big(1+\frac{\xi k}{\beta}\Big)^{-1/\xi}
\sim K\Big(\frac{\beta}{\xi}\Big)^{1/\xi}k^{-1/\xi},\qquad k\to\infty.
\]
Thus $\overline F_Y(k)=k^{-1/\xi}L(k)$ with $L(k)\to K(\beta/\xi)^{1/\xi}$ slowly varying, so $\overline F_Y\in RV_{-1/\xi}$. BY argument of \citet{hitz_davis_samorodnitsky_2024} yields $F_Y\in\mathrm{MDA}_\xi$, $\xi>0$.
\end{proof}
}

\answerRrR{
\section{Theoretical results of EGPD and DEGPD Model 3}
Let $Z\geq 0$ be a continuous random variable. For
\[
\mathcal{G}(\nu;\kappa)= 
\begin{cases}
0, & \nu<0,\\[6pt]
\dfrac{D_{\{(1/2 -\zeta)\nu-\zeta\}}(\kappa, \kappa)- D_{\zeta}(\kappa, \kappa)}{D_{1/2}(\kappa, \kappa)-D_{\zeta}(\kappa, \kappa)}, & 0 \leq \nu \leq 1,\\[8pt]
1, & \nu>1,
\end{cases}
\]
 the defined $\mathcal{G}(\nu)$ is the cdf of truncated Beta distribution. The detail can be see in the main denouement. The CDF of the EGPD Model 3 is written as
\begin{equation}\label{cdf-tn}
H_Z(z;\kappa, \beta, \xi) = \mathcal{G}\big(G(z, \beta, \xi);\kappa\big),
\end{equation}
where $G(., \beta, \xi)$ is the CDF of GPD. The parameter $\kappa>0$ is a shape parameter controlling the lower tail behavior, $\xi>0$ is a shape parameter
controlling the upper tail behavior, and $\sigma>0$ an overall scale parameter affecting the whole
distribution from low to high quantiles.
}

\answerRrR{
\begin{theorem}[Regular variation of beta-mapped EGPD]
Let $Z\sim H_Z(\cdot;\kappa, \beta, \xi)$ with $\kappa>0,\ \beta>0,\ \xi>0$, where
\[
H_Z(z;\kappa, \beta, \xi) \;=\; \mathcal{G}\big(G(z);\kappa\big),
\]
\[
\mathcal{G}(u;\kappa)= 
\begin{cases}
0, & u<0,\\[6pt]
\dfrac{D_{(1/2 -\zeta)u+\zeta}(\kappa, \kappa)- D_{\zeta}(\kappa, \kappa)}{D_{1/2}(\kappa, \kappa)-D_{\zeta}(\kappa, \kappa)}, & 0 \leq u \leq 1,\\[8pt]
1, & u>1,
\end{cases}
\]
and $G(z)=1-(1+\xi z/\beta)^{-1/\xi}$ is the GPD CDF. Assume $0\le \zeta<1/2$. Then:
\begin{enumerate}
\item The survival function satisfies
\[
\overline H_Z(z)\sim C\, \omega(z), \qquad z\to\infty,
\]
where $\omega(z)=(1+\xi z/\beta)^{-1/\xi}$ and
\[
C=\frac{f_{\mathrm{Beta}(\kappa,\kappa)}(1/2)\,(1/2-\zeta)}{D_{1/2}(\kappa,\kappa)-D_{\zeta}(\kappa,\kappa)}.
\]
\item As $z\to\infty$,
\[
\omega(z)\sim\left(\frac{\beta}{\xi}\right)^{1/\xi} z^{-1/\xi}.
\]
Therefore
\[
\overline H_Z(z)\sim C\left(\frac{\beta}{\xi}\right)^{1/\xi} z^{-1/\xi},\qquad z\to\infty.
\]
\item Hence $\overline H_Z\in RV_{-1/\xi}$ and $H_Z$ belongs to the Fr\'echet maximum domain of attraction with index $\xi$, i.e., 
$H_Z\in \mathrm{MDA}_\xi.$
\end{enumerate}
\end{theorem}
}
\answerRrR{
\begin{proof} We initially set
\[
u:=G(z)=1-\Big(1+\frac{\xi z}{\beta}\Big)^{-1/\xi},\qquad
\varepsilon:=1-u=\Big(1+\frac{\xi z}{\beta}\Big)^{-1/\xi}=\omega(z).
\]
Put $\delta:=\tfrac12-\zeta>0$. Denote by $D_x(\kappa,\kappa)$ the Beta$(\kappa,\kappa)$ CDF at $x$ and by $f_{\mathrm{Beta}(\kappa,\kappa)}(x)$ its density:
\[
f_{\mathrm{Beta}(\kappa,\kappa)}(x)=\frac{x^{\kappa-1}(1-x)^{\kappa-1}}{B(\kappa,\kappa)}.
\]
As $z\to\infty$ we have $\varepsilon\to0$ and $u=1-\varepsilon\to1$. The Beta argument is
\[
(1/2-\zeta)u+\zeta=(1/2-\zeta)(1-\varepsilon)+\zeta
=1/2-(1/2-\zeta)\varepsilon=1/2-\delta\varepsilon,
\]
so it approaches $1/2$ as $u\to1$.
By Taylor expansion of the Beta CDF about $x=1/2$ (the density is smooth),
\[
D_{1/2-\delta\varepsilon}(\kappa,\kappa)
= D_{1/2}(\kappa,\kappa) - f_{\mathrm{Beta}}(1/2)\,\delta\varepsilon + O(\varepsilon^2).
\]
The numerator of $\mathcal G$ is
\begin{align*}
D_{(1/2-\zeta)u+\zeta}(\kappa,\kappa)-D_{\zeta}(\kappa,\kappa)
&= D_{1/2-\delta\varepsilon}(\kappa,\kappa)-D_{\zeta}(\kappa,\kappa)\\
&= \big[D_{1/2}(\kappa,\kappa)-D_{\zeta}(\kappa,\kappa)\big] - f_{\mathrm{Beta}}(1/2)\,\delta\varepsilon + O(\varepsilon^2).
\end{align*}
Dividing by the denominator $D_{1/2}(\kappa,\kappa)-D_{\zeta}(\kappa,\kappa)$ yields
\[
\mathcal G(u;\kappa) = 1 - \frac{f_{\mathrm{Beta}}(1/2)\,\delta}{D_{1/2}(\kappa,\kappa)-D_{\zeta}(\kappa,\kappa)}\,\varepsilon + O(\varepsilon^2).
\]
Therefore,
\[
\overline H_Z(z)=1-H_Z(z)=\frac{f_{\mathrm{Beta}}(1/2)\,\delta}{D_{1/2}(\kappa,\kappa)-D_{\zeta}(\kappa,\kappa)}\,\varepsilon + O(\varepsilon^2),
\]
so $\overline H_Z(z)\sim C\,\omega(z)$ with
$
C=\frac{f_{\mathrm{Beta}}(1/2)\,\delta}{D_{1/2}(\kappa,\kappa)-D_{\zeta}(\kappa,\kappa)}.
$
\\
The tail $\omega(z)=(1+\xi z/\beta)^{-1/\xi}$ satisfies
\[
\omega(z)\sim\Big(\frac{\beta}{\xi}\Big)^{1/\xi} z^{-1/\xi},\qquad z\to\infty.
\]
Combining with the survival function  gives
\[
\overline H_Z(z)\sim C\Big(\frac{\beta}{\xi}\Big)^{1/\xi} z^{-1/\xi},
\]
hence $\overline H_Z\in RV_{-1/\xi}$ and $H_Z\in\mathrm{MDA}_\xi$.
\end{proof}
}

\begin{figure}{}
\centering
\subfloat[DEGPD ($M_1$)]{%
\resizebox*{15cm}{!}{\includegraphics{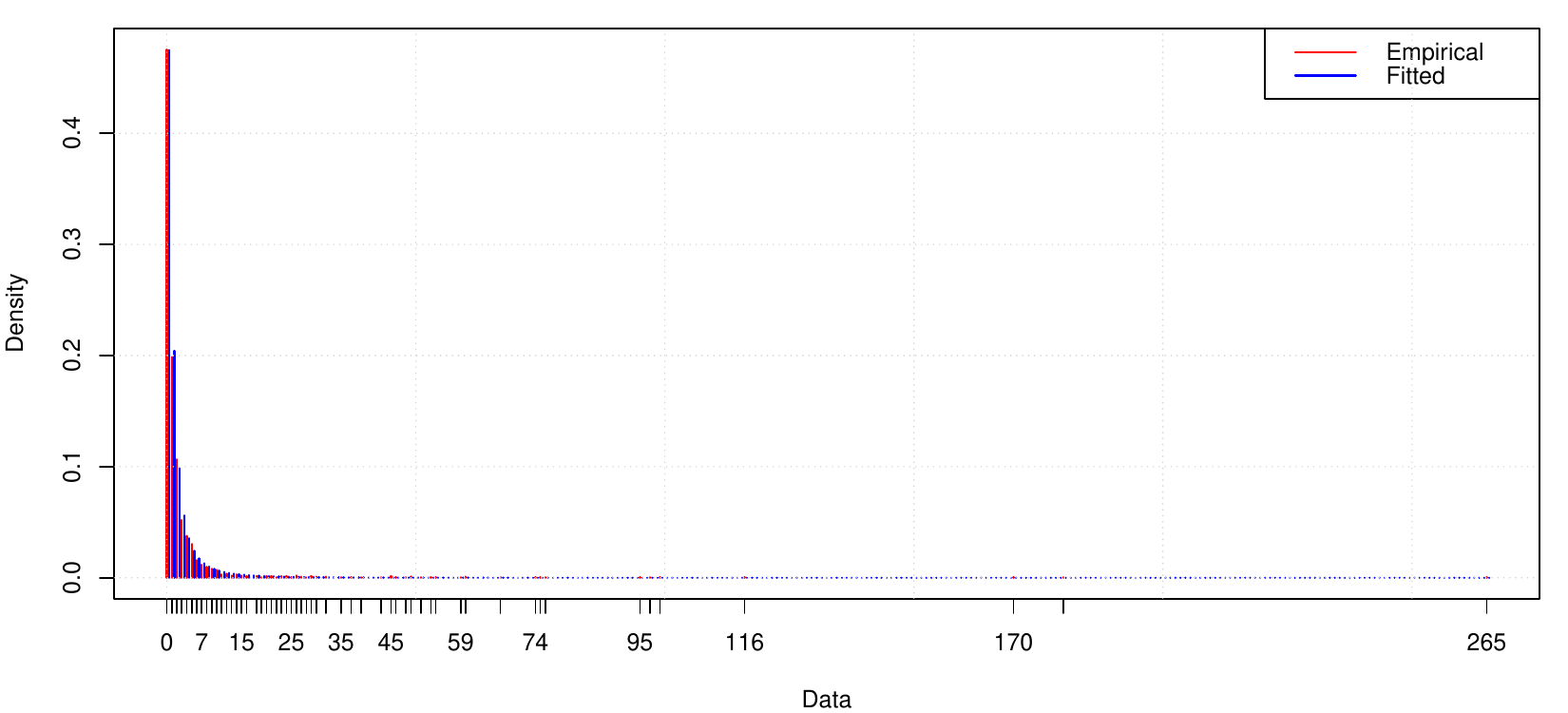}}}\hspace{5pt}\\
\subfloat[DEGPD ($M_2$)]{%
\resizebox*{15cm}{!}{\includegraphics{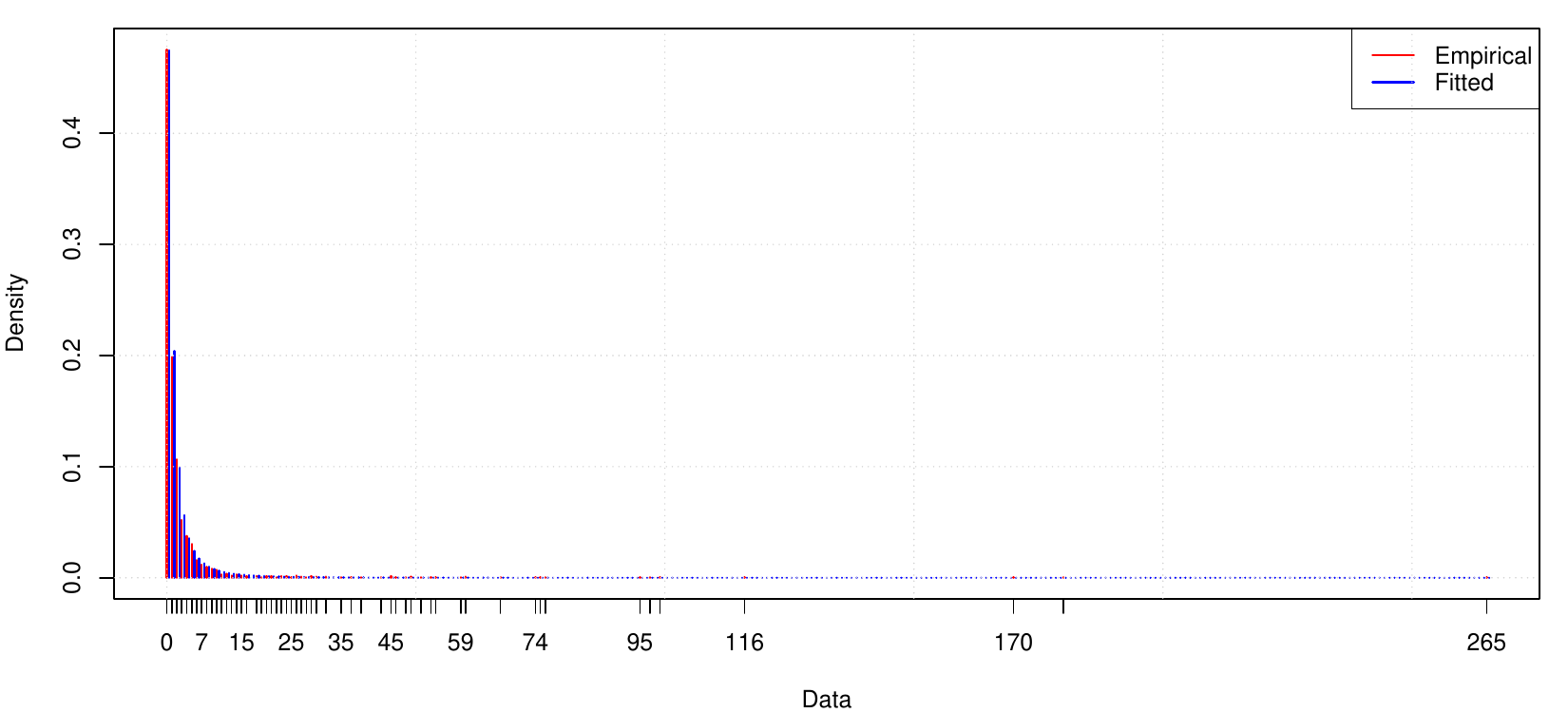}}}\\
\subfloat[DEGPD ($M_3$)]{%
\resizebox*{15cm}{!}{\includegraphics{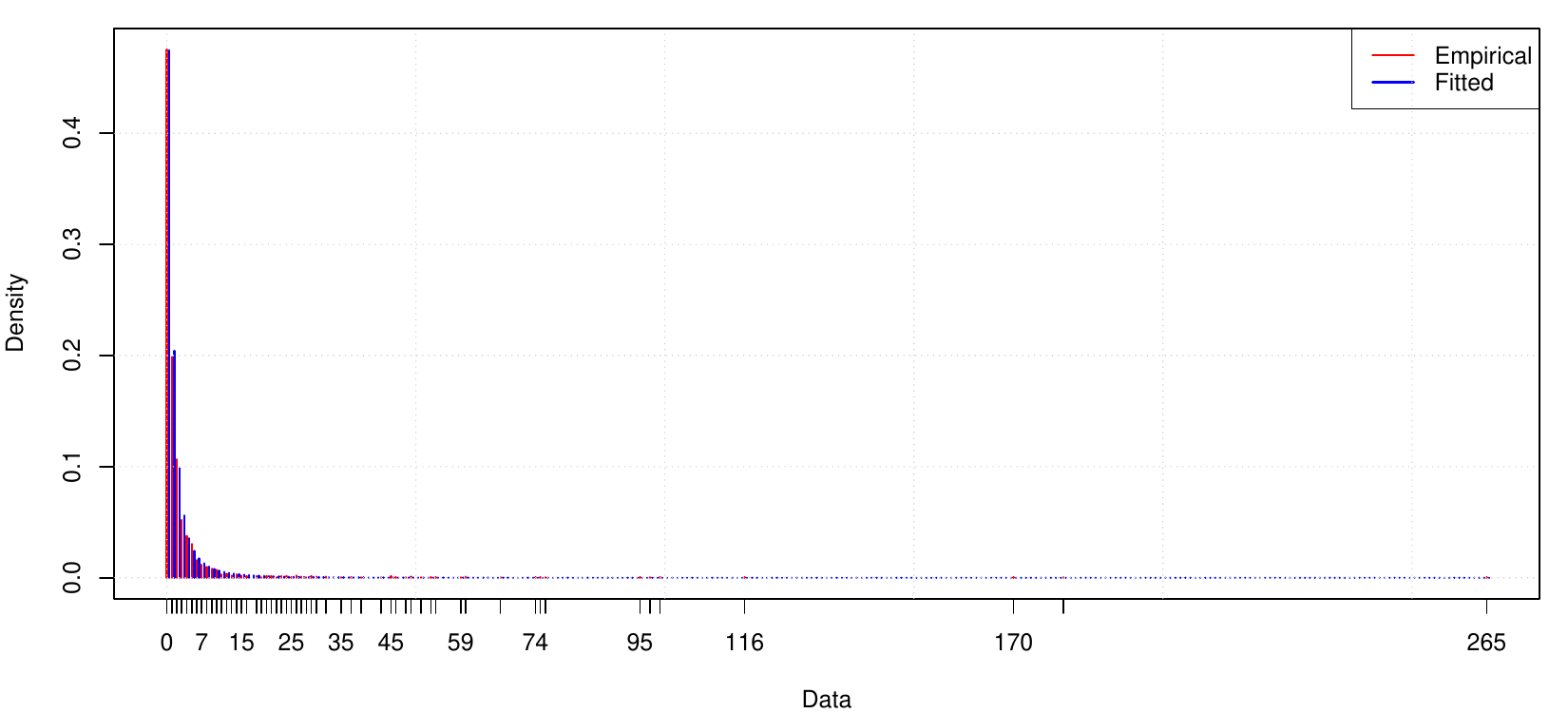}}}
 \caption{Fitted and empirical distribution plot for DEGPD models $M_1$–$M_3$.}
	\label{fig:DEGPD-sup}
\end{figure}

\answerRrR{
\begin{theorem}[Discrete beta-mapped EGPD]
Let $Z\sim H_Z(\cdot;\kappa,\beta,\xi)$ satisfy the assumptions and conclusion of the
previous theorem (so that $\overline H_Z(z)\sim C\,\omega(z)$ with
$\omega(z)=(1+\xi z/\beta)^{-1/\xi}$ and $C>0$). Define the integer-valued
random variable $Y=\lfloor Z\rfloor$. Then:
\begin{enumerate}
\item The discrete survival function satisfies
\[
\overline F_Y(k)=\Pr(Y\ge k)\sim C\left(\frac{\beta}{\xi}\right)^{1/\xi} k^{-1/\xi},
\qquad k\to\infty.
\]
\item Hence $\overline F_Y\in RV_{-1/\xi}$ and $F_Y$ belongs to the discrete Fr\'echet
maximum domain of attraction with index $\xi$, i.e. $F_Y\in\mathrm{MDA}_\xi$.
\end{enumerate}
\end{theorem}
}
\answerRrR{
\begin{proof}
By the definition of the floor,
\[
\{Y\ge k\}=\{\lfloor Z\rfloor\ge k\}=\{Z\ge k\},
\]
so for each integer $k\ge0$,
\[
\overline F_Y(k)=\Pr(Y\ge k)=\Pr(\lfloor Z\rfloor\ge k)=\overline H_Z(k).
\]
From the continuous case theorem, $\overline H_Z(z) \sim C(\beta/\xi)^{1/\xi} z^{-1/\xi}$.
Since $\overline F_Y(k) = \Pr(Z \ge k) = \overline F_Z(k)$, we have:
\[
\overline F_Y(k) \sim C\left(\frac{\beta}{\xi}\right)^{1/\xi} (k)^{-1/\xi}
\]
Therefore $\overline F_Y(k) = k^{-1/\xi} L(k)$ where $L(k) \to C(\beta/\xi)^{1/\xi}$ is slowly varying, so $\overline F_Y \in RV_{-1/\xi}$.
\\
For discrete distributions with $\overline F_Y(k) \in RV_{-1/\xi}$, $\xi > 0$. Following the argument of~\citet{hitz_davis_samorodnitsky_2024}, we implies that $F_Y$ belongs to the 
 discrete maximum domain of attraction with shape parameter $\xi>0$. 
This completes the proof.
\end{proof}
}
\answerRrR{
\section{Real application results}
We fitted the DEGPD models ($M_1$–$M_3$) to the Automobile Insurance Company Complaint Rankings (AICCR) for New York City. Details of the dataset are provided in the main document. Figure~\ref{fig:DEGPD-sup} shows the fitted and empirical density plots, illustrating that all three models closely capture the empirical distribution of the observed data.}

\begin{figure}
\centering
\subfloat[ZIDEGPD ($M_1$)]{%
\resizebox*{7cm}{!}{\includegraphics{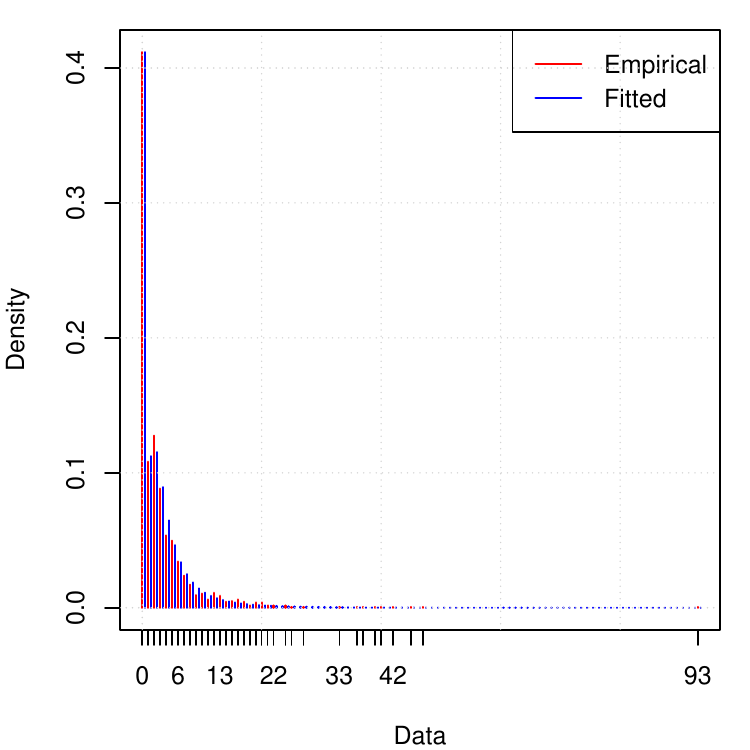}}}\hspace{5pt}
\subfloat[ZIDEGPD ($M_2$)]{%
\resizebox*{7cm}{!}{\includegraphics{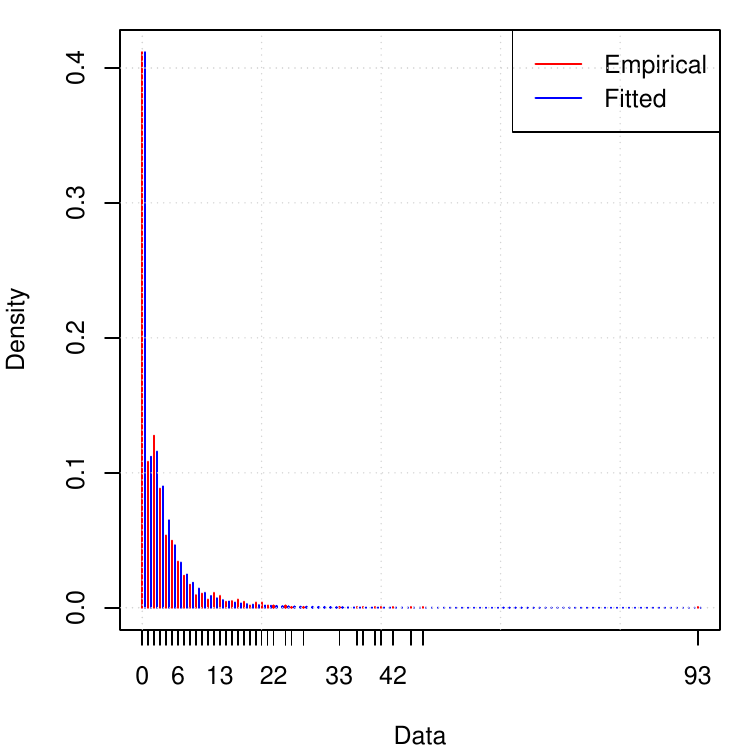}}}\\
\subfloat[ZIDEGPD ($M_3$)]{%
\resizebox*{7cm}{!}{\includegraphics{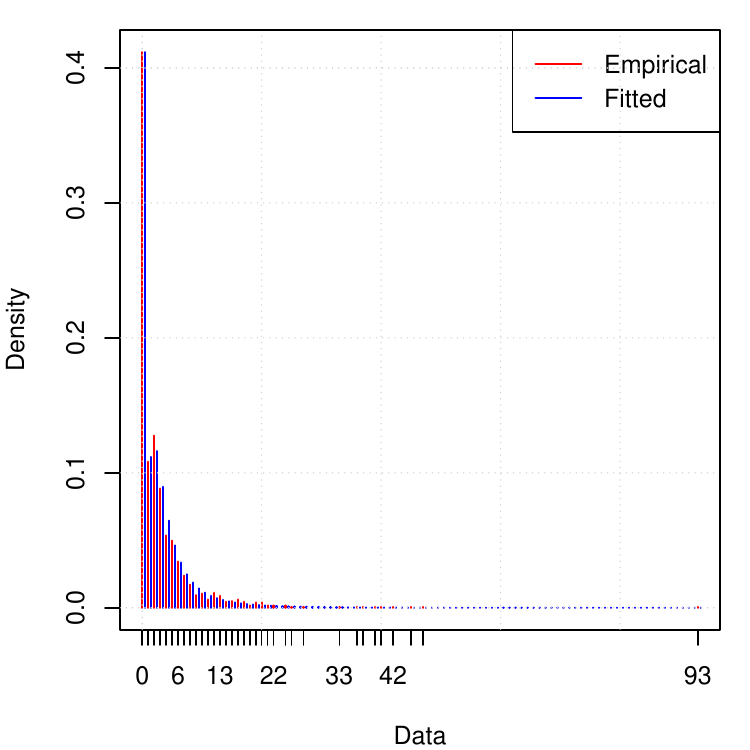}}}
 \caption{Fitted and empirical distribution plot for ZIDEGPD models $M_1$–$M_3$.}
	\label{fig:ZIGEPD-sup}
\end{figure}

\begin{figure}
\centering
\subfloat[DEGPD ($M_1$)]{%
\resizebox*{7cm}{!}{\includegraphics{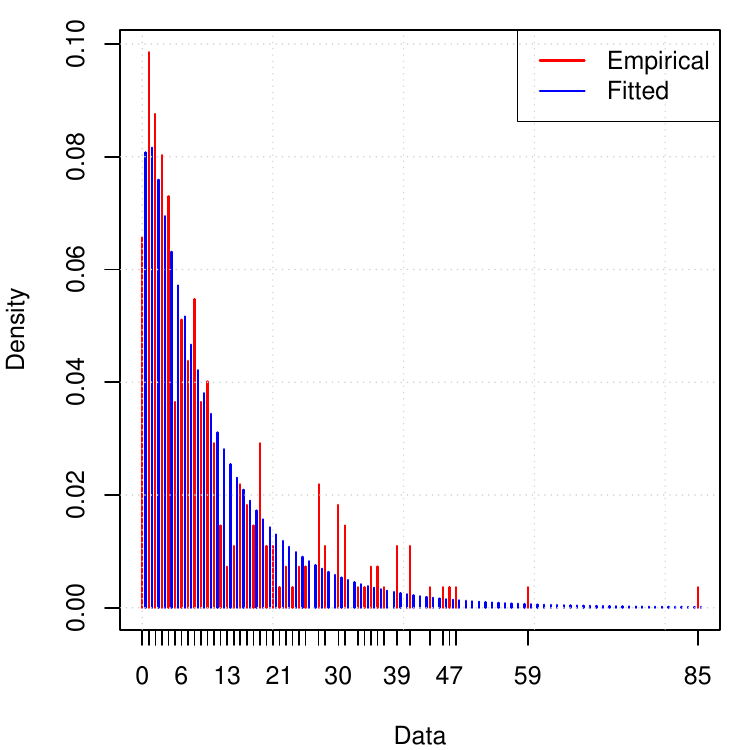}}}\hspace{5pt}
\subfloat[DEGPD ($M_2$)]{%
\resizebox*{7cm}{!}{\includegraphics{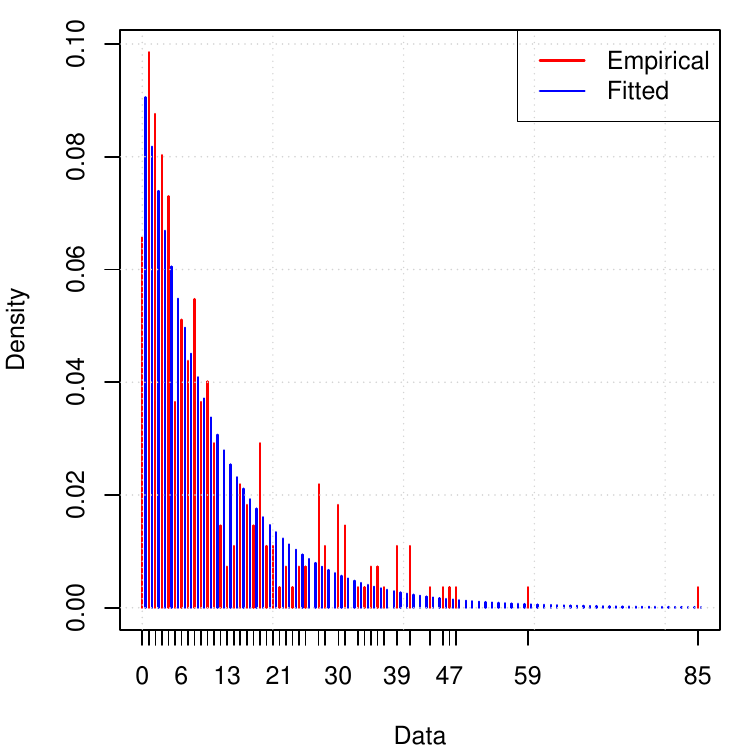}}}\\
\subfloat[DEGPD ($M_3$)]{%
\resizebox*{7cm}{!}{\includegraphics{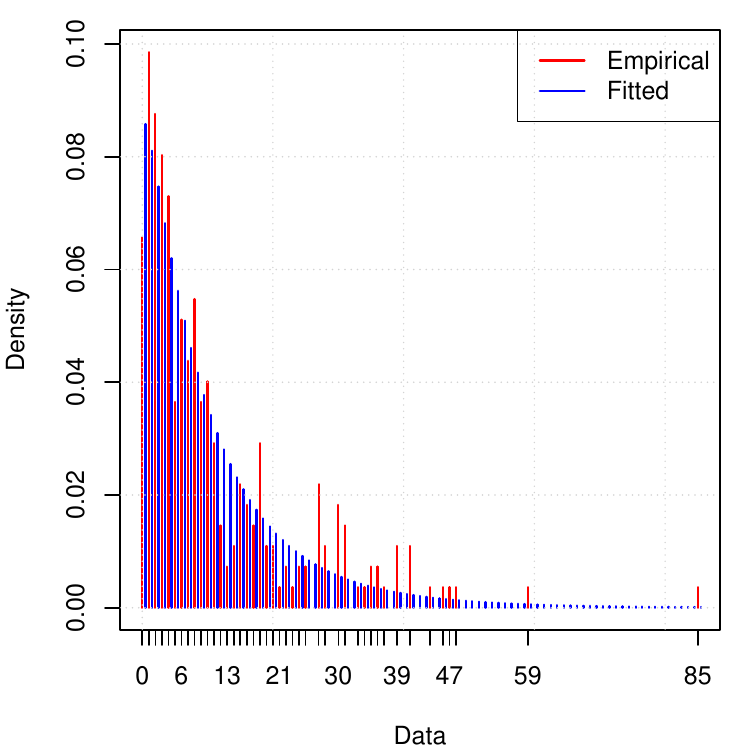}}}
 \caption{Fitted and empirical distribution plot for DEGPD models $M_1$–$M_3$ at threshold $u=3$.}
	\label{fig:DEGPD-sup-thresh}
\end{figure}

\answerRrR{We fitted the ZIDEGPD models ($M_1$–$M_3$) to the dataset of doctors' hospital visits, a classic example of zero-inflated data with a heavy tail. Figure~\ref{fig:ZIGEPD-sup} shows the fitted and empirical density plots, indicating that all three models closely match the observed distribution.
\\
The DEGPD models were also applied to a third dataset on betting and gaming offenses in New South Wales, Australia, using a low threshold of 3. Figure~\ref{fig:DEGPD-sup-thresh} presents the fitted and empirical density plots, showing similar performance across models, with $M_2$ slightly better aligned with the empirical distribution in the lower tail.
}

\end{document}